\documentclass[twocolumn,prd,superscriptaddress,floatfix,preprintnumbers,letterpaper]{revtex4-1}

\usepackage{array}
\usepackage{dcolumn}
\usepackage{bm}
\usepackage[colorlinks, pdfborder={0 0 0}, plainpages=false]{hyperref}
\usepackage{graphicx}
\usepackage{xspace}
\usepackage[usenames,dvipsnames]{color}
\usepackage{amssymb}
\usepackage[normalem]{ulem} 
\usepackage{lineno}
\usepackage{letltxmacro}
\usepackage{amsfonts}
\usepackage{amsmath}
\usepackage{amssymb}
\usepackage{bm}
\usepackage{dcolumn}
\usepackage{epsfig}
\usepackage{graphicx}
\usepackage{graphics}
\usepackage[latin1]{inputenc}
\usepackage{latexsym}
\usepackage{rotating}
\usepackage{hyperref}
\usepackage[usenames]{color}
\usepackage{float}
\usepackage{xspace} 
\usepackage{mathrsfs}
\usepackage{subfigure}
\usepackage{multirow}
\usepackage{booktabs}
\usepackage{soul}

\usepackage{colortbl}
\usepackage{etoolbox}
\usepackage[table]{xcolor}

\def\prd{{\it Phys. Rev.} D~}

\def\apjl{{\it Astrophys. J.} Lett~}
\def\apj{{\it Astrophys. J.}}

\def\PRD{{\it Phys. Rev.} D~}

\def\mnras{{\it MNRAS~}}

\def\lesssim{\mathrel{\hbox{\rlap{\hbox{\lower4pt\hbox{$\sim$}}}\hbox{$<$}}}}
\def\gtrsim{\mathrel{\hbox{\rlap{\hbox{\lower4pt\hbox{$\sim$}}}\hbox{$>$}}}}
\def\alt{\mathrel{\hbox{\rlap{\hbox{\lower4pt\hbox{$\sim$}}}\hbox{$<$}}}}
\def\agt{\mathrel{\hbox{\rlap{\hbox{\lower4pt\hbox{$\sim$}}}\hbox{$>$}}}}

\usepackage{mathtools}
\DeclarePairedDelimiter\abs{\lvert}{\rvert}

\def\gta{\ifmmode {\mathbin{\lower 3pt\hbox   
    {$\,\rlap{\raise 5pt\hbox{$\char'076$}}\mathchar"7218\,$}}}
    \else {${\mathbin{\lower 3pt\hbox
    {$\rlap{\raise 5pt\hbox{$\char'076$}}\mathchar"7218\,$}}}
    $}\fi}
\def\lta{\ifmmode {\,\mathbin{\lower 3pt\hbox   
    {$\,\rlap{\raise 5pt\hbox{$\char'074$}}\mathchar"7218\,$}}}
    \else {${\mathbin{\lower 3pt\hbox
    {$\rlap{\raise 5pt\hbox{$\char'074$}}\mathchar"7218\,$}}}
    $}\fi}

\newcommand{\beq}{\begin{equation}}
\newcommand{\eeq}{\end{equation}}
\newcommand{\bea}{\begin{eqnarray}}
\newcommand{\eea}{\end{eqnarray}}

\definecolor{darkperiwinkle}{RGB}{102, 102, 128}

\newcommand{\NCSA}{\affiliation{NCSA, University of Illinois at Urbana-Champaign, Urbana, Illinois 61801, USA}}
\newcommand{\ANCSA}{\affiliation{Department of Astronomy, University of Illinois at Urbana-Champaign, Urbana, Illinois 61801, USA}}
\newcommand{\PNCSA}{\affiliation{Department of Physics, University of Illinois at Urbana-Champaign, Urbana, Illinois 61801, USA}}

\definecolor{light-gray}{gray}{0.9}

\begin{document}

\title{Characterization of numerical relativity waveforms \\ of eccentric binary black hole mergers}

\author{Sarah Habib}\NCSA\PNCSA
\author{E. A. Huerta}\NCSA\ANCSA

\date{\today}

\begin{abstract}
\noindent We introduce a method to quantify the initial eccentricity, gravitational wave frequency, and mean anomaly of numerical relativity simulations that describe non-spinning black holes on moderately eccentric orbits. We demonstrate that this method provides a robust characterization of eccentric binary black hole mergers with mass-ratios \(q\leq10\) and eccentricities \(e_0\leq0.2\) fifteen cycles before merger. We quantify the circularization rate of a variety of eccentric numerical relativity waveforms introduced in~\cite{ecc_catalog} by computing overlaps with their quasi-circular counterparts, finding that \(50M\) before merger they attain overlaps \({\cal{O}}\geq0.99\), furnishing evidence for the circularization of moderately eccentric binary black hole mergers with mass-ratios \(q\leq10\). We also quantify the importance of including higher-order waveform modes for the characterization of eccentric binary black hole mergers. Using two types of numerical waveforms, one that includes \((\ell, \, \abs{m})= \{(2,\,2),\, (2,\,1),\, (3,\,3),\, (3,\,2), \, (3,\,1),\, (4,\,4),\, (4,\,3),\, (4,\,2),\,(4,\,1)\}\) and one that only includes  the \(\ell=\abs{m}=2\) mode, we find that the overlap between these two classes of waveforms is as low as \({\cal{O}}=0.89\) for \(q=10\) eccentric binary black hole mergers, underscoring the need to include higher-order waveform modes for the description of these gravitational wave sources. We discuss the implications of these findings for future source modeling and gravitational wave detection efforts.

\end{abstract}

\pacs{Valid PACS appear here}
\maketitle

\section{Introduction}
\label{sec:intro}

Numerical relativity (NR) has played a key role in the discovery and interpretation of 
gravitational wave (GW) 
observations~\cite{DI:2016,secondBBH:2016,thirddetection,fourth:2017,GW170608,o1o2catalog,bnsdet:2017}.
As the LIGO~\cite{DII:2016,LSC:2015} and Virgo~\cite{Virgo:2015} observatories 
continue to probe the GW spectrum, NR will provide key insights to infer the properties of 
binary black hole (BBH) mergers whose GWs exhibit strong spin-precession, or high-order 
waveform modes~\cite{Chu:2016CQG,Mroue:2013,Kumar:2016dhh,NRI:2016,NRGW1509142,Lange:2017PhRvD,huerta:2018PhRvD,Adam:2018arXiv,Hinder:2010,Shawn_R_2019,Huerta:2017a,sper:2008PhRvDS,hinder:2017a,sum:2009CQGra,kotake:2013CRPhy,Tanja:2018a,Fou:2018arX,rad:2018arXiv18}.
These improved studies will be key to infer the formation channels of these objects, and 
to ascertain whether they are accurately described by general relativity. 

BBH systems that form through massive stellar evolution in the field of galaxies are expected to 
enter the frequency band of LIGO-type detectors with nearly circular orbits~\cite{Postnov2014}. Under this assumption, 
NR groups have produced thousands of NR waveforms to get insights into the physics of these 
GW sources~\cite{Mroue:2013,SXS_Catalog_2019,RIT_Catalog_II,RIT_Catalog_I,GaTech_Catalog}. 
These NR waveforms have been used to calibrate semi-analytical models~\cite{Bohe:2016gbl,Tara:2014,eobnon,gsve,cot_buo_2019,london:170800404L,husacv:2016PhRvD,khan:2016PhRvD,Huerta:2017a}, produce 
fast interpolators using gaussian process regression~\cite{huerta:2018PhRvD,varma:2019PhRvD}, and surrogate models~\cite{blackman:2015,blackman_su_II_2017} to 
inform the development of signal-processing tools for GW searches~\cite{2016CQGra..33u5004U,GstLAL_pap,geodf:2017b,geodf:2017a,dgNIPS,hshen:2017,DL_Workshop:2018,shen_icasp:2019,geodf:2017c,wei:2019W,Adam:2018arXiv}, and more recently to directly infer 
the astrophysical properties of BBH mergers through GW observations~\cite{NRI:2016,Kumar_w_NR_BBH,Lange:2017PhRvD}.

BBHs may also form in dense stellar environments, such as globular clusters and galactic nuclei~\cite{sam:2017ApJ,Samsing:2014,sam:2018PhRvD3014S,Leigh:2018MNRAS,ssm:2017,ssm:2018,samsing:2018ApJ140S,lisa:2018b,Huerta:2009,sam:2017ApJ84636S,sam:2018MNRAS1548S,Huerta:2015a,sam:2019MNRAS30S,sam:2018PhRvD3014S,sam:2018MNRAS5436S,Huerta:2014,Anton:2014,samdor:2018MNRAS5445S,samdorII:2018MNRAS4775D,samdor:2018arXiv64S,samjoh:2018Z,rocarl:2018PhRvDR,kremerjoh:2018aK,lopez:2018L,hoang:2017APJ,gon:2017,hpoang:2017,lisa:2018a,MikKoc:2012,Naoz:2013,gondkoc:2018G,antonras:2016ApJ7A,Huerta:2013a,arcakoc:2018A,takkoc:2018ApJT,gondkoc:2018ApJ5G,antoni:2018A,Anto:2015arXiv,lisa:2018b,lisa:2018a,samsing_venu_2019,lopez_sam_2018,Rod_Car_2018PhRDR,Zevin_Sam_2019ApJZ}. Electromagnetic 
observations provide evidence for their existence in galactic clusters, and in the center of 
the Milky Way~\cite{galcen:2018,Sippel:2013,Strader:2012,ssm:2018}. 
This increasing body of observational evidence has sparked the interest of the 
community to better understand these sources. This program includes GW source modeling, 
formation channels and merger rates, and astrophysically motivated scenarios where GW observations 
may be used to confirm or rule out the existence of BBHs in dense stellar environments. On this latter point, 
it is now widely accepted that the measurement of orbital eccentricity through GW observations would be the cleanest signature to furnish evidence for the existence of compact binary populations in dense stellar environments. It is for this reason that the GW source modeling community is 
sharpening its analytical and numerical 
tools to infer the imprints of orbital eccentricity in GW searches.~\cite{East:2013,cao:2017,Hinderer:2017,Huerta:2017a,Huerta:2009,Huerta:2015a,huerta:2018PhRvD,Huerta:2014,moore:2018fde,lou:2016arXiv,lou:2017CQG,yunes-eccentric-2009}.  

An accurate description of the physics of eccentric BBHs throughout the late-inspiral, 
merger and ringdown requires NR~\cite{ihh:2008PhRvD,Hinder:2010,east:2012a,east:2012,Gold:2012PG,Gold:2013,East:2015PRDa,East:2016PhRvD,Radice:2016MNRAS,2016arXiv161107531T,lewis:2017CQG,ecc_catalog}.
Once the data products of NR simulations are post-processed, and NR waveforms are extracted~\cite{johnson:2017}, 
it is necessary to characterize them, i.e., we need 
to quantify the eccentricity and other orbital parameters that uniquely identify them. 
One can address this task using a variety of methods. If BBHs are on nearly quasi-circular orbits,
then one could use the approach introduced in~\cite{initial_data_RIT:2017CQG}, which combines 
information about the orbital separation of the BHs, and waveform phase and amplitude of the 
Weyl scalar \(\psi_4\). This methodology only includes \({\cal{O}}(e)\) corrections to measure orbital eccentricity, 
which limits its applicability to characterize moderately eccentric 
BBH mergers. Some other methods try to infer orbital eccentricity based on the trajectories of the 
BHs in the NR simulation, which is not a sound approach given that these trajectories are gauge-dependent.

We circumvent the aforementioned limitations by introducing a gauge-invariant method that 
characterizes a NR waveform by comparing to a large array of \texttt{ENIGMA} waveforms~\cite{huerta:2018PhRvD}. 
The inspiral-merger-ringdown \texttt{ENIGMA} model consists of an 
inspiral evolution that encodes higher-order post-Newtonian (PN) corrections to the motion of compact sources 
on eccentric orbits, combined with self-force and BH perturbation corrections. This approach ensures that the 
dynamics of quasi-circular and moderately eccentric BBHs are accurately described. Assuming that 
moderately eccentric systems circularize prior to merger, we attach a stand-alone quasi-circular merger 
waveform to the inspiral evolution. The merger waveforms are produced using a Gaussian Process Emulator~\cite{Moore160125}
that is trained with NR waveforms describing quasi-circular BH mergers. 

In summary, our goal is to constrain the PN parameters that produce the 
optimal overlap between \texttt{ENIGMA} waveforms and their NR counterparts. 
This study is timely and relevant if we are to use NR waveforms to properly characterize future 
observations of eccentric BBH mergers. Once we showcase the application of this method, 
we also quantify the 
circularization of eccentric BBHs near merger, 
and quantify the impact of higher-order waveform modes in the morphology of eccentric NR waveforms. 

This paper is organized as follows. In Section~\ref{sec:method} we describe how we adapted the \texttt{ENIGMA}
model to characterize eccentric NR waveforms, and present results of this method in Section~\ref{sec:dem}. In Section~\ref{sec:circular} we study the circularization of 
moderately eccentric BBH mergers that retain eccentricity a few cycles before merger. 
We use overlap calculations to quantify the impact of higher-order waveform modes on the morphology of eccentric BBH mergers in Section~\ref{sec:hom}. We summarize our findings and future directions of 
work in Section~\ref{sec:end}.


\section{Methods}
\label{sec:method}

We measure the orbital eccentricity of NR waveforms using the \(\ell=\abs{m}=2\) mode. The rationale for this 
choice is that we are characterizing NR waveforms using the \texttt{ENIGMA} waveform model, which only includes the \(\ell=\abs{m}=2\) mode.

We explored a variety of gauge-invariant objects to directly compare NR and \texttt{ENIGMA} waveforms, 
and found that the dimensionless object \(M\omega\), where \(\omega\) is the mean orbital frequency, and 
\(M\) stands for the total mass of the BBH, provides a robust approach to capture the signatures of eccentricity. 
To compute \(M\omega\) using our NR waveforms, we first extract the waveform modes \(h_{\ell,\,m}\) at future null infinity using the open source software \texttt{POWER}~\cite{johnson:2017}. As mentioned above, since we will be comparing \(M\omega\) with a waveform model that only includes the \(\ell=\abs{m}=2\) mode, we take only the waveform mode \(h^{(\ell=\abs{m}=2)}(t)\) to compute \(M\omega\) using the following relations:

\begin{eqnarray}
\label{eq:polar}
h^{(\ell=\abs{m}=2)}(t) &=& h_{+} - i h_{\times}\,,\\
\label{eq:momega}
M\omega &=& M\dot{\phi}^{(\ell=\abs{m}=2)} = \frac{1}{2}\frac{\dot{h}_{+}\, h_{\times} - h_{+}\,\dot{h}_{\times}}{h^2_{+}+h^2_{\times}}\,,
\end{eqnarray}

\noindent where \(\dot{\phi}^{(\ell=\abs{m}=2)}\) is the unwrapped GW phase of \(h^{(\ell=\abs{m}=2)}(t)\). On the other hand, the quantity  \(M\omega\) is one of the building blocks of the \texttt{ENIGMA} model, 
which includes tail, tails-of-tails and non-linear memory corrections at 1.5PN, 2.5PN and 3PN 
order~\cite{Huerta:2017a,huerta:2018PhRvD}. To be self-contained, it is worth mentioning that the mean orbital frequency, \(\omega\), is related to the mean motion, \(n\), through the relation \(\omega = {\cal{K}} n\), where \({\cal{K}}\) is the periastron precession. Furthermore, the mean anomaly, \(l\), is related to the mean motion, \(n\), through the relation \(M\dot{l} = M n\)~\cite{Blanchet:2006}.

\subsection{Optimization algorithm}
\begin{figure}
\centerline{
\includegraphics[width=0.54\linewidth]{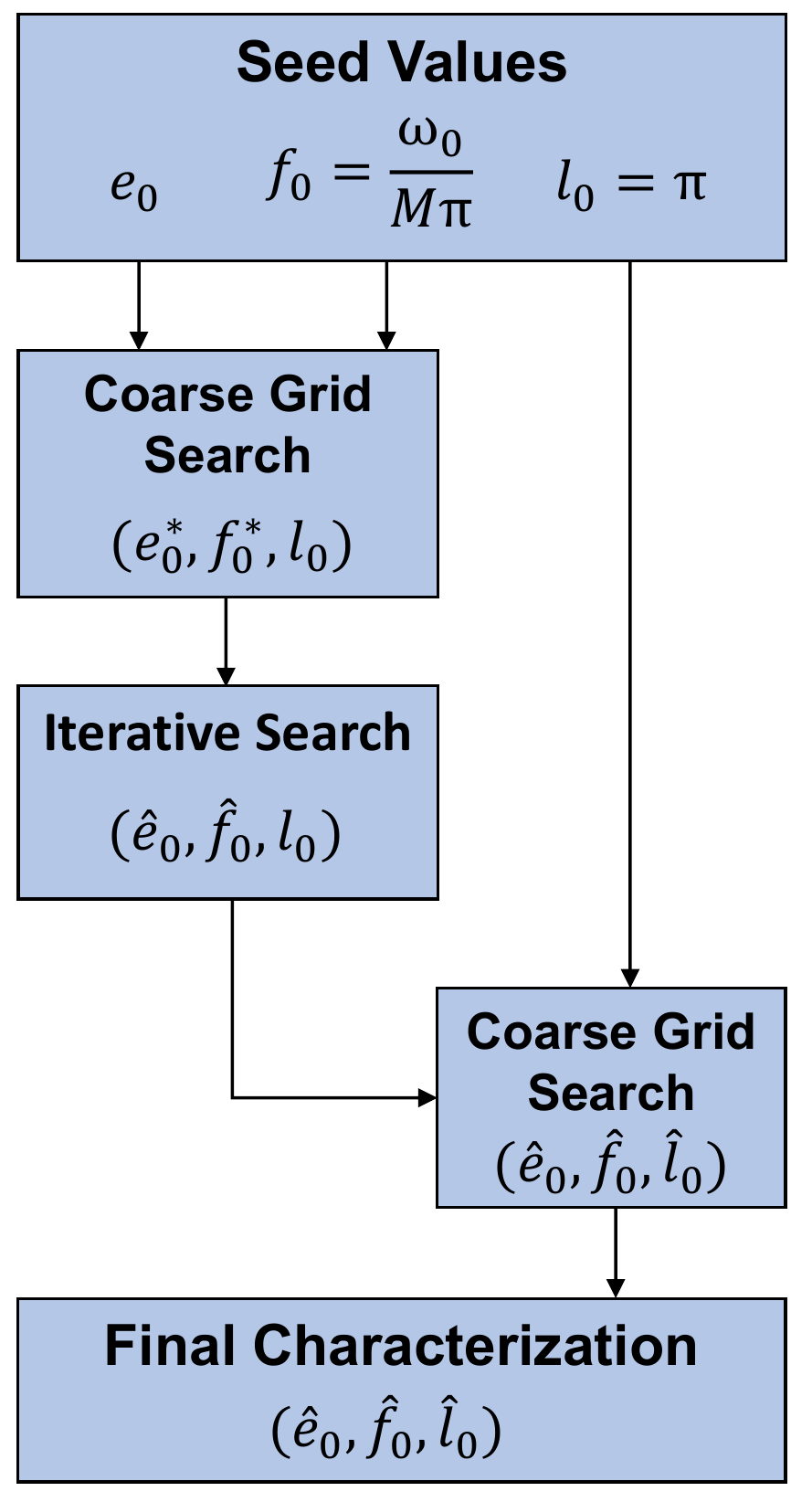}
}
\caption{A schematic of the algorithm used to characterize NR waveforms. First, \(f_0\) and \(e_0\) are roughly estimated using a grid search determining the optimal coordinate pair \((e^*_0,f^*_0)\). An iterative search is then performed on \(f^*_0\) and \(e^*_0\) to increase precision. With these refined \(\hat{f}_0\) and \(\hat{e}_0\), a second grid search is used to find the optimal \(\hat{l}_0\). The output is the optimal triple \((\hat{e}_0,\hat{f}_0,\hat{l}_0)\).} 
 \label{fig:workflow}
 \end{figure}

In order to characterize an NR waveform, we compare it to \texttt{ENIGMA} waveforms generated algorithmically. Initial conditions of the \texttt{ENIGMA} simulation are varied until the \texttt{ENIGMA} time evolution of \(M\omega\) agrees with the NR evolution to a specified degree of accuracy. The \texttt{ENIGMA} parameters of concern are initial orbital eccentricity \(e_0\), initial GW frequency \(f_0\), and mean anomaly \(l_0\). This method is schematically described in Figure~\ref{fig:workflow}. Following~\cite{GopakumarandK:2006}, we parameterize the orbital eccentricity, \(e\), in the \texttt{ENIGMA} model using the PN time eccentricity parameter, i.e., \(e^{\textrm{PN}}_t\rightarrow e\).

\begin{figure}
\centerline{
\includegraphics[width=\linewidth]{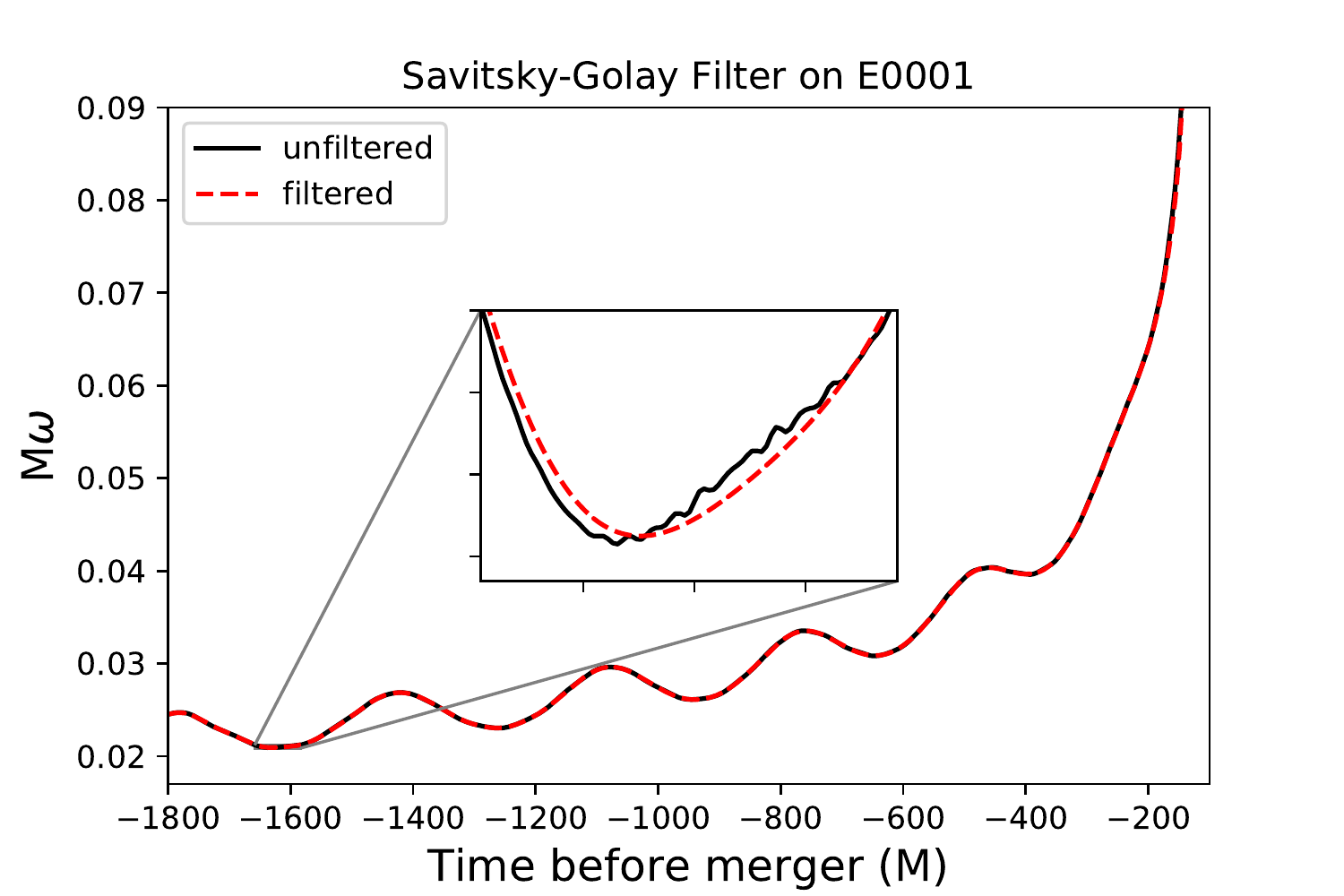}
}
\caption{Removal of high frequency noise from \(M\omega\) using a Savitsky-Golay filter. Notice that this filtering scheme does not remove the true eccentricity features from the time-series data.} 
 \label{fig:filter}
 \end{figure}

The NR waveform is first preprocessed by cutting off initial noise (junk radiation). We then apply a Savitsky-Golay filter to remove high frequency noise from the \(M\omega\) time-series data. We have used the \texttt{Python} implementation of the Savitsky-Golay filter, which is extensively used to remove high frequency noise from data while preserving the original features of the signal better than other types of filtering approaches, such as moving averages techniques~\cite{filter_sg}. Applying this filter also allows for more accurate detection of local extrema in the signal. Figure~\ref{fig:filter} presents the input and output time-series data when we use this filter to remove high-frequency noise from the \(M\omega\) time-series data.

From a modeling perspective, we know that both eccentricity and gravitational wave frequency are coupled (this is the reason why many semi-analytical approaches have been explored in recent years using the stationary phase approximation~\cite{moore_2019M,moore:2018fde,Huerta:2014,yunes-eccentric-2009}). In view of this observation, we optimize \((e_0, f_0)\) simultaneously. On the other hand, we know that the net effect of varying \(l_0\) is to introduce what amounts to a minor shift in the position of the peaks of the orbital frequency \(M\omega\). To reduce computational time, we perform a separate, coarser search for \(l_0\) instead of optimizing it simultaneously with the other parameters. Since we know experimentally that values close to \(l_0\sim\pi\) will produce \texttt{ENIGMA} waveforms that tend to be aligned with their NR counterparts, we start the optimization with this seeded value for \(l_0\) and then do minor refinement to this value later on.

Since input parameters are required to produce \texttt{ENIGMA} waveforms, seed values are initialized for \(f_0\) and \(l_0\). We provide an informed guess of GW frequency using the relation \(\omega_0/\left(M \pi\right)\), where \(\omega(t_0)=\omega_0\) and \(t_0\) is the time at which the NR waveform is free from junk radiation. Mean anomaly is initialized to \(\pi\), a value manually determined to be optimal through verification of a few individually sampled NR waveforms. Orbital eccentricity does not require a seed value since the range of possible values is consistent for all catalogued waveforms, so the following grid search directly samples eccentricity from a predefined range.

\begin{figure*}[htp]
\centerline{
\includegraphics[width=0.54\linewidth]{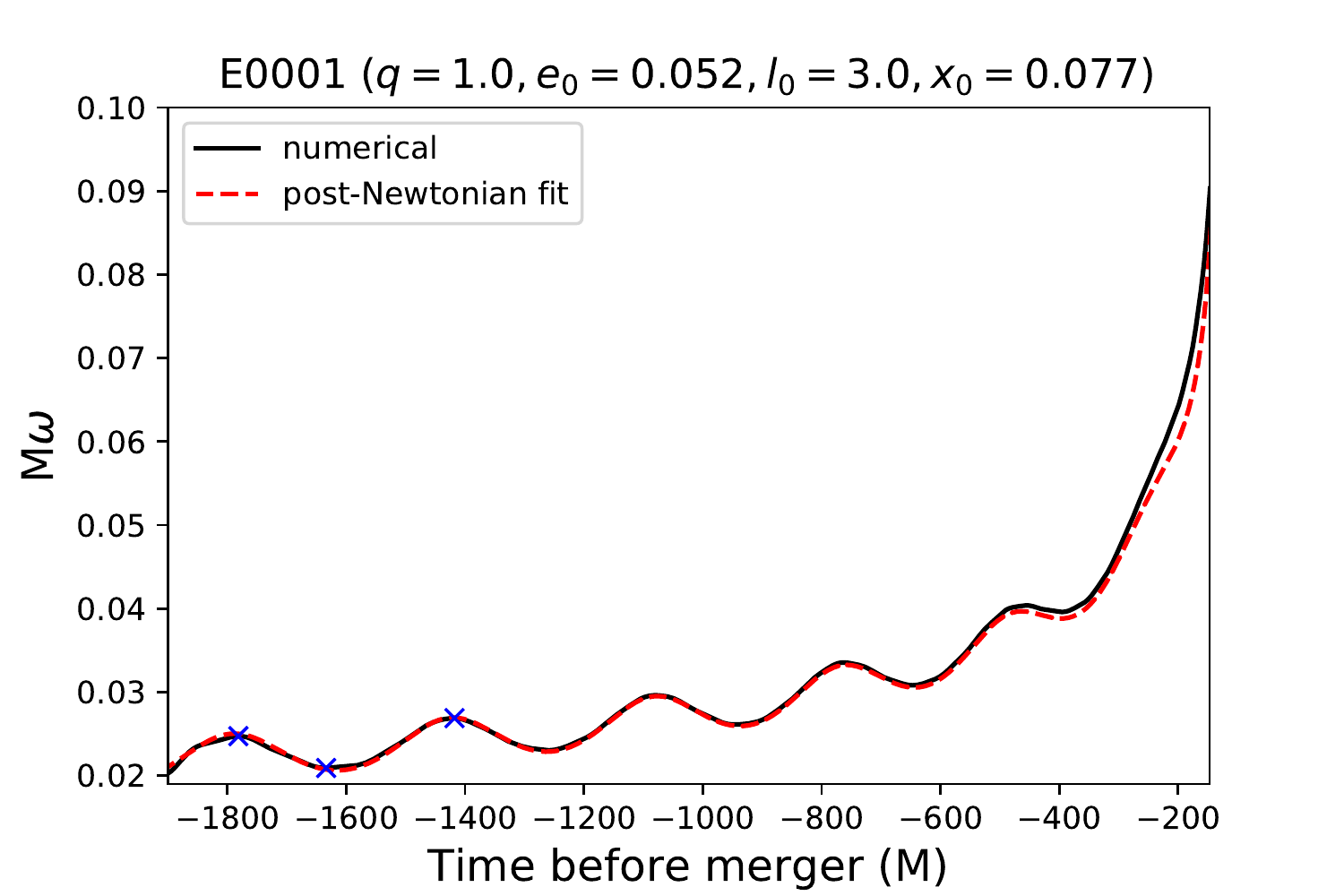}\hspace{-1.5em}%
\includegraphics[width=0.54\linewidth]{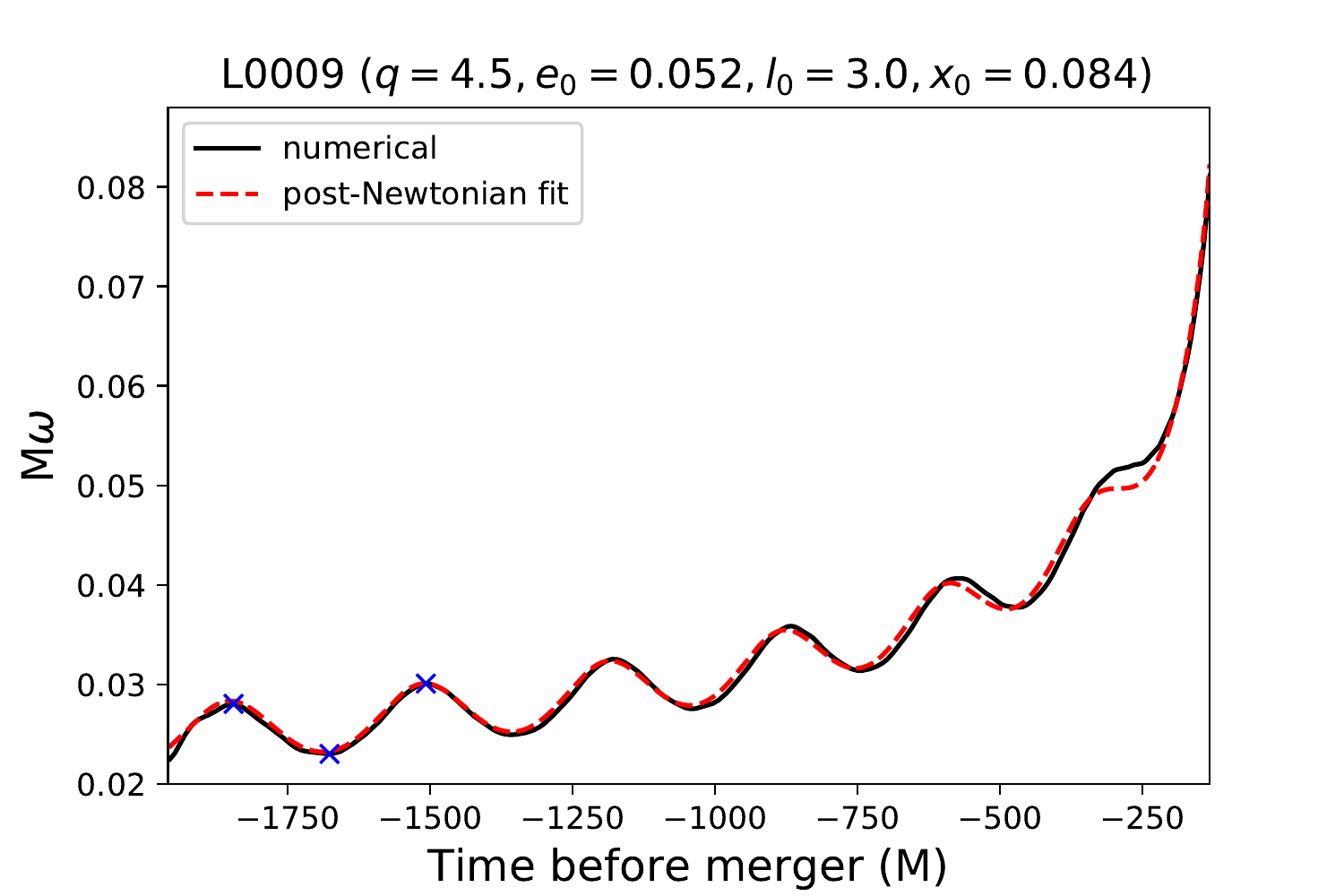}\hspace{-1.5em}%
}
\centerline{
\includegraphics[width=0.54\linewidth]{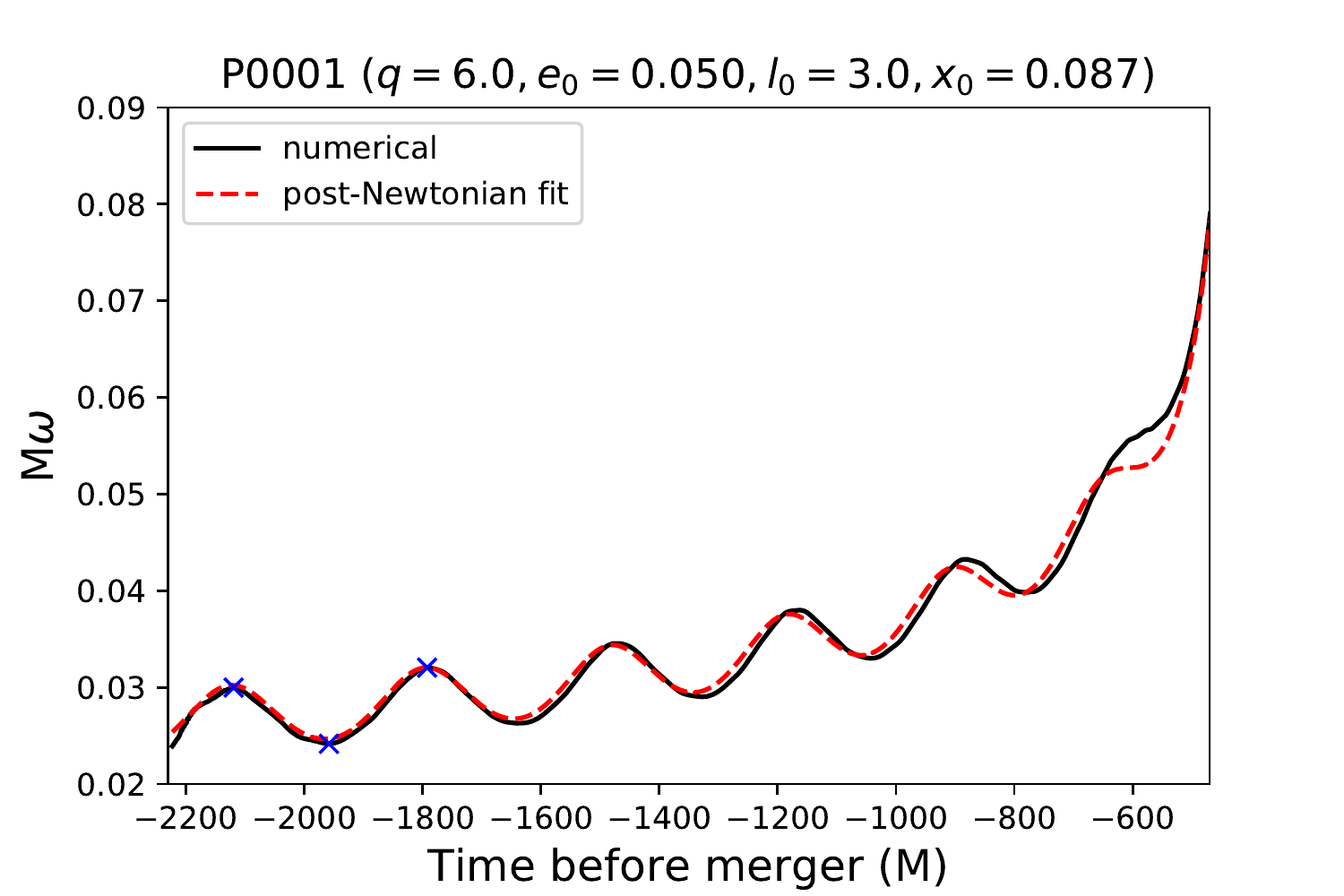}\hspace{-1.5em}
\includegraphics[width=0.54\linewidth]{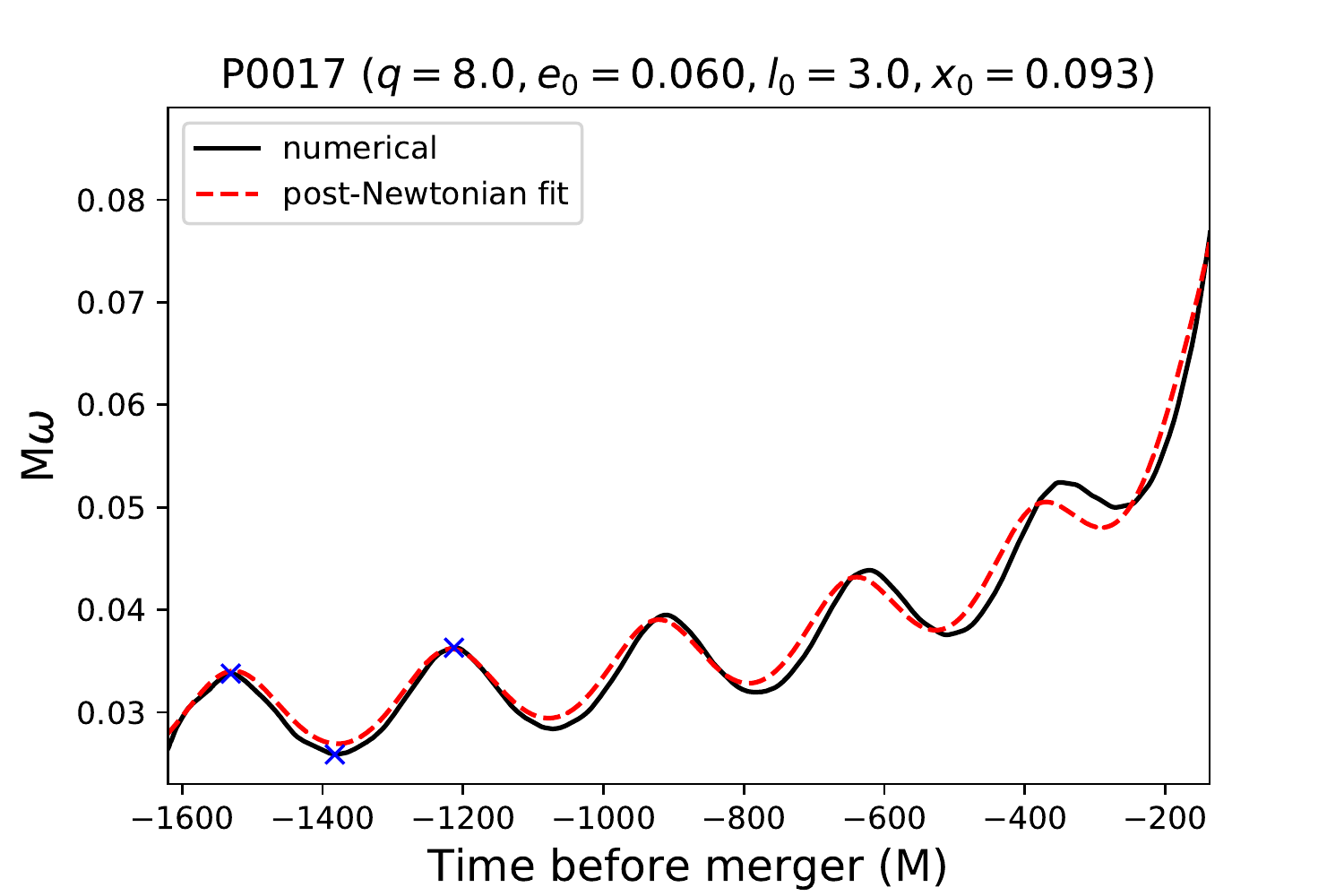}\hspace{-1.5em}
}
\centerline{
\includegraphics[width=0.54\linewidth]{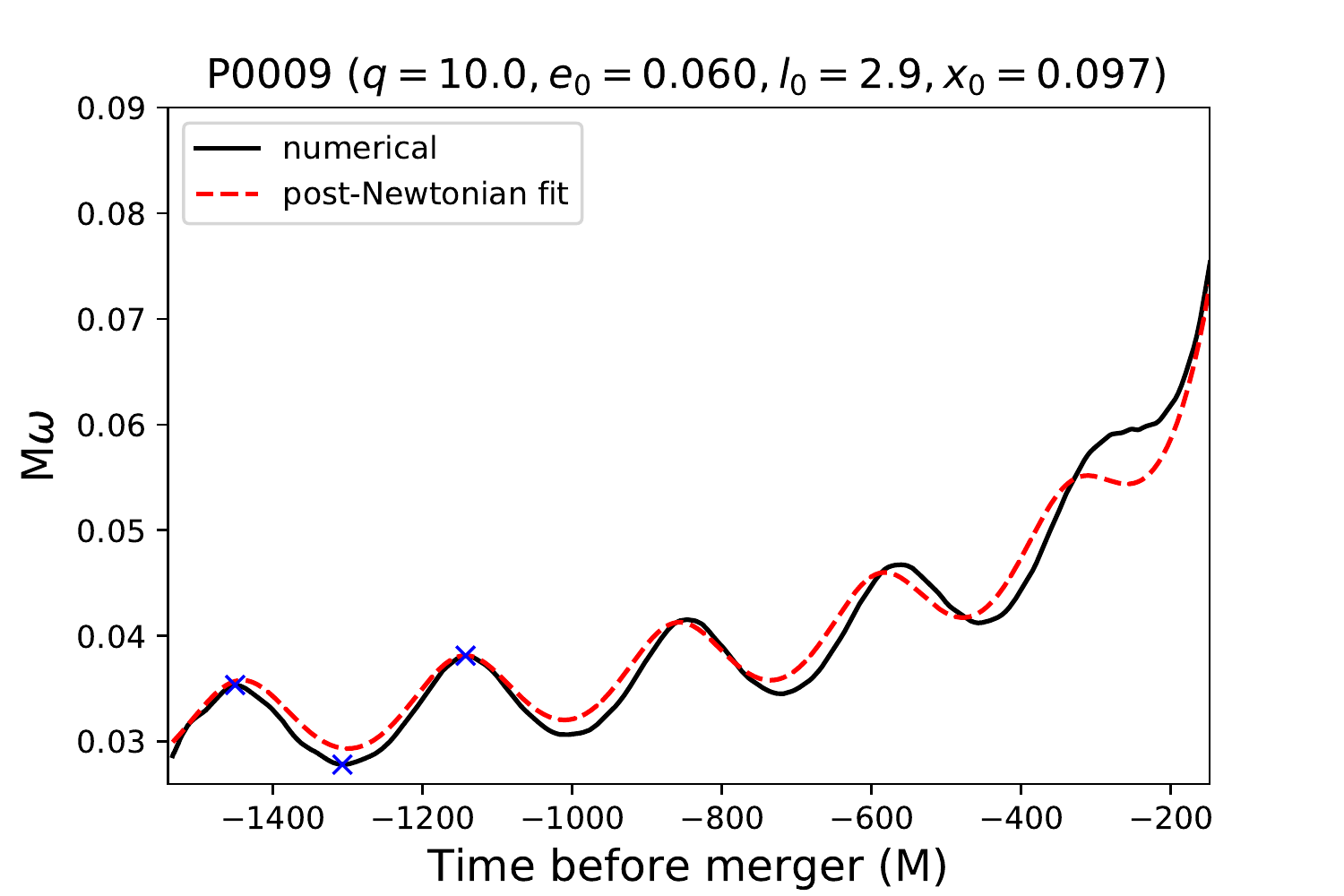}
}
\caption{Comparison between the dimensionless orbital frequency, \(M\omega\), of numerical relativity simulations and their corresponding post-Newtonian counterparts. Each post-Newtonian evolution was constructed using the \texttt{ENIGMA} model and optimal values for the triplet \((\hat{e}_0,\hat{f}_0,\hat{l}_0)\) provided by the algorithm described in Section~\ref{sec:method}. The extrema used in search cost functions are marked.} 
 \label{fig:optimization}
 \end{figure*}

The algorithm starts with a grid search in the 2D parameter space of \((f_0,e_0)\), and iteratively refines the resulting parameter guess. To generate the grid, we densely sample the frequency range \( f\in[f_0-5\textrm{Hz},\,f_0+5\textrm{Hz}] \) and the eccentricity range \(e_0\in[0.1,\,0.3]\). For each coordinate pair, an \texttt{ENIGMA} \(M\omega\) is produced using the specified \((f_0,\,e_0)\) values and the seeded \(l_0\). The resulting \(M\omega\) time evolution is then compared to that of the original NR waveform. Parameters are chosen that minimize two cost functions. Specifically, the optimal coordinate pair is that which minimizes the two cost functions \(|t^{FC}_\texttt{{ENIGMA}} - t^{FC}_{NR}|\) and \(|A^{*}_{\texttt{ENIGMA}} - A^{*}_{NR}|\), where \(t^{FC}\) is the time duration of the first orbital cycle (time between two consecutive maxima, as shown in Fig.~\ref{fig:optimization}) and \(A^*\) is the amplitude (difference in \(M\omega\) from first maximum to first minimum, as shown in Fig.~\ref{fig:optimization}) respectively. Because the eccentricity is most easily measurable early on in the waveform time evolution, only the first orbital cycle is considered in these cost functions. Throughout this grid search, \(l_0\) is held constant.

\noindent After completing the grid search, the chosen \((f_0,\,e_0)\) parameters are further refined iteratively using a hill-climbing approach. In this stage, the initial GW frequency and orbital eccentricity are independently varied stepwise. In order to avoid the hill-climbing problem of finding a solution at a local minimum of the cost function, the initial guesses of \((f_0,\, e_0)\) are randomly sampled within the range confined by the grid search, and the greedy search is repeated until parameters are found that vary by $\leq 1\%$. This ``greedy" search is run for a preset number of iterations. At each iteration, an \texttt{ENIGMA} waveform is generated using the current \((f_0,\,e_0)\) guesses and the seeded \(l_0\). Properties of the \(M\omega\) evolution are then compared between the \texttt{ENIGMA} and NR models to determine how to increment \(e_0\) and \(f_0\). Initial eccentricity is evaluated depending on the amplitude of the first orbital cycle $A^{*}$, calculated again as the difference in $M\omega$ from first maximum to first minimum. If $A^*_{\texttt{ENIGMA}} > A^*_{\textrm{NR}}$, then the current $e_0$ guess is too large, so $e_0$ is decreased; vice versa is true for $A^*_{\texttt{ENIGMA}} < A^*_{\textrm{NR}}$. Initial frequency is varied based on the time \(t^*\) each simulation reaches the threshold frequency, which corresponds to the time at which a quasi-circular waveform would be attached in the \texttt{ENIGMA} model. If \(t^*_{\texttt{ENIGMA}} - t^*_{\textrm{NR}} > 0\), the current \(f_0\) guess is too low, so \(f_0\) is increased; conversely, if \(t^*_{\texttt{ENIGMA}} - t^*_{\textrm{NR}} < 0\), the current \(f_0\) guess is decreased. To increase parameter precision, increment step size is one degree of precision higher than that of the grid search resolution. Using as input seeds the values found for \((f_0,\,e_0)\), we constrain \(l_0\) using an additional grid search. This second grid search finds the \(l_0\) (sampled from the range \(l_0\in[0,\,2\pi)\)) minimizing the cost function \(|t^M_{\textrm{ENIGMA}} - t^M_{\textrm{NR}}|\), where \(t^M\) is the difference in time of first orbital cycle occurrence between the \texttt{ENIGMA} and NR waveforms.

Upon finding the optimal values \((\hat{f}_0,\,\hat{e}_0,\,\hat{\ell}_0)\), we recast the initial parameter into one that is more commonly used to describe NR waveforms, i.e., \(\hat{x}_0 = \left(M\omega_0\right)^{2/3}\). In Figure~\ref{fig:optimization} we present results of this optimization procedure for a sample of NR waveforms. Note that the \(M\omega\) time-series datasets presented therein have already been cleaned from high-frequency noise using a Savitsky-Golay filter, as shown in Figure~\ref{fig:filter}. We used this method to characterize the 89 NR waveforms presented in~\cite{ecc_catalog}. The properties of the BBHs considered in this study are summarized in Table~\ref{sims}.

 \begin{table}
\caption{The table presents the mass-ratio, \(q\), and the measured values of initial eccentricity, mean anomaly and dimensionless orbital frequency, \((\hat{e}_0,\, \hat{l}_0,\, \hat{x}_0)\), at a time, \(t_0\), when the numerical relativity simulations are free from junk radiation.}
		\footnotesize
		\begin{center}
                        \setlength{\tabcolsep}{10pt} 
			\begin{tabular}{c c c c c c}
				\hline 
				Simulation&$q$ & $\hat{e}_0$ & $\hat{l}_0$ & $\hat{x}_0$ \\ 
				\hline
				E0001	&	1.0	&	0.052	&	3.0	&	0.0770	\\
				K0006	&	4.0	&	0.068	&	3.0	&	0.0826	\\
				L0009	&	4.5	&	0.052	&	3.0	&	0.0839	\\
				L0016	&	5.0	&	0.140	&	2.9	&	0.0862	\\
				P0001	&	6.0	&	0.050	&	3.0	&	0.0867	\\
				P0017	&	8.0	&	0.060	&	3.0	&	0.0927	\\
				P0006	&	8.0	&	0.080	&	2.9	&	0.0931	\\
				P0007	&	8.0	&	0.100	&	2.9	&	0.0926	\\
				P0009	&    10.0	&	0.060	&	2.9	&	0.0971	\\
				P0022	&    10.0	&	0.080	&	2.9	&	0.0979	\\
				P0023	&    10.0	&	0.120	&	2.9	&	0.0968	\\
				P0024	&    10.0	&	0.180	&	3.0	&	0.0957	\\
				\hline 
			\end{tabular}
		\end{center}
	\label{sims}
	\end{table}
	\normalsize

\section{Characterization of eccentric numerical relativity waveforms}
\label{sec:dem}
 
\begin{figure*}[htp]
\centerline{
\includegraphics[width=0.5\linewidth]{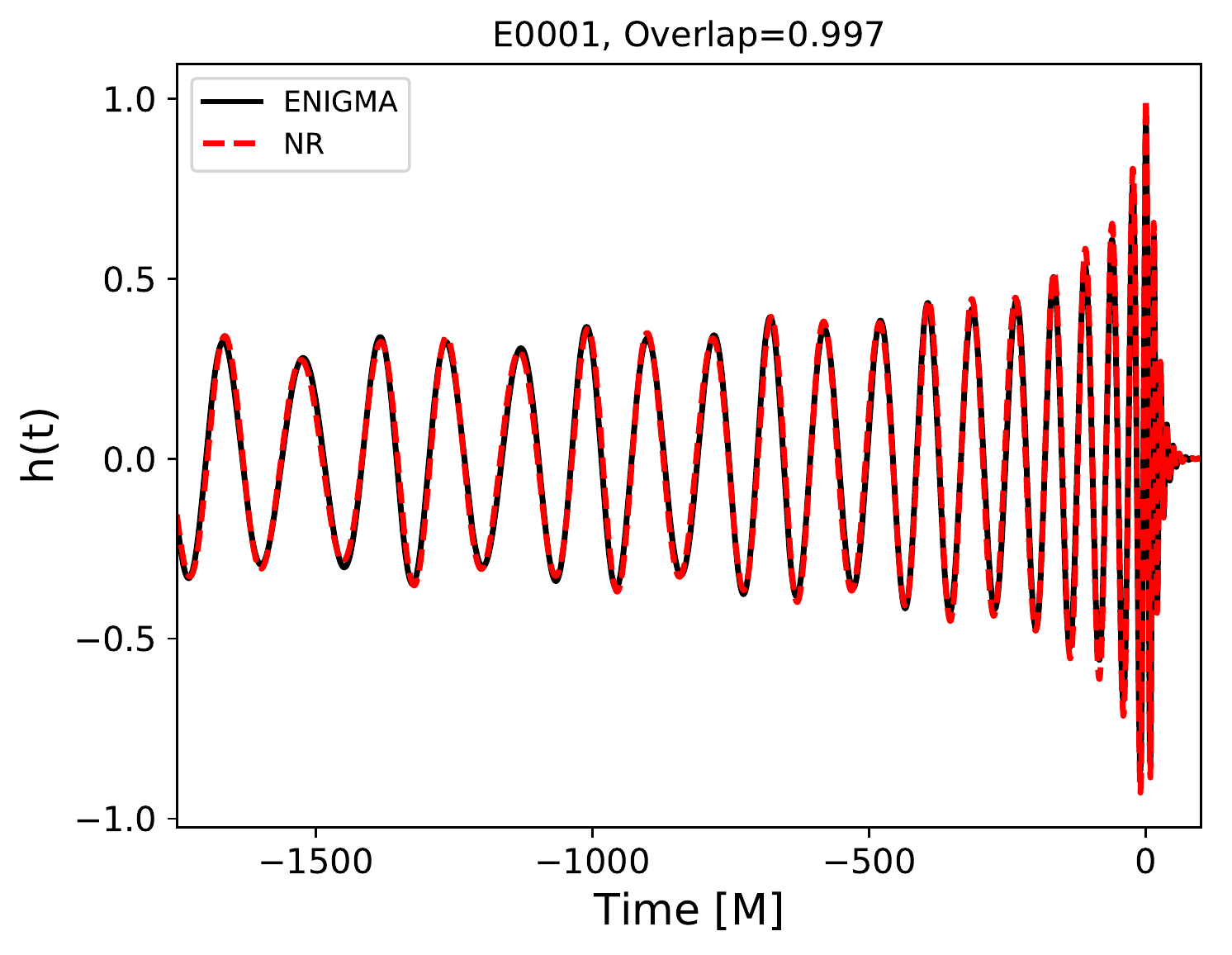}
\includegraphics[width=0.5\linewidth]{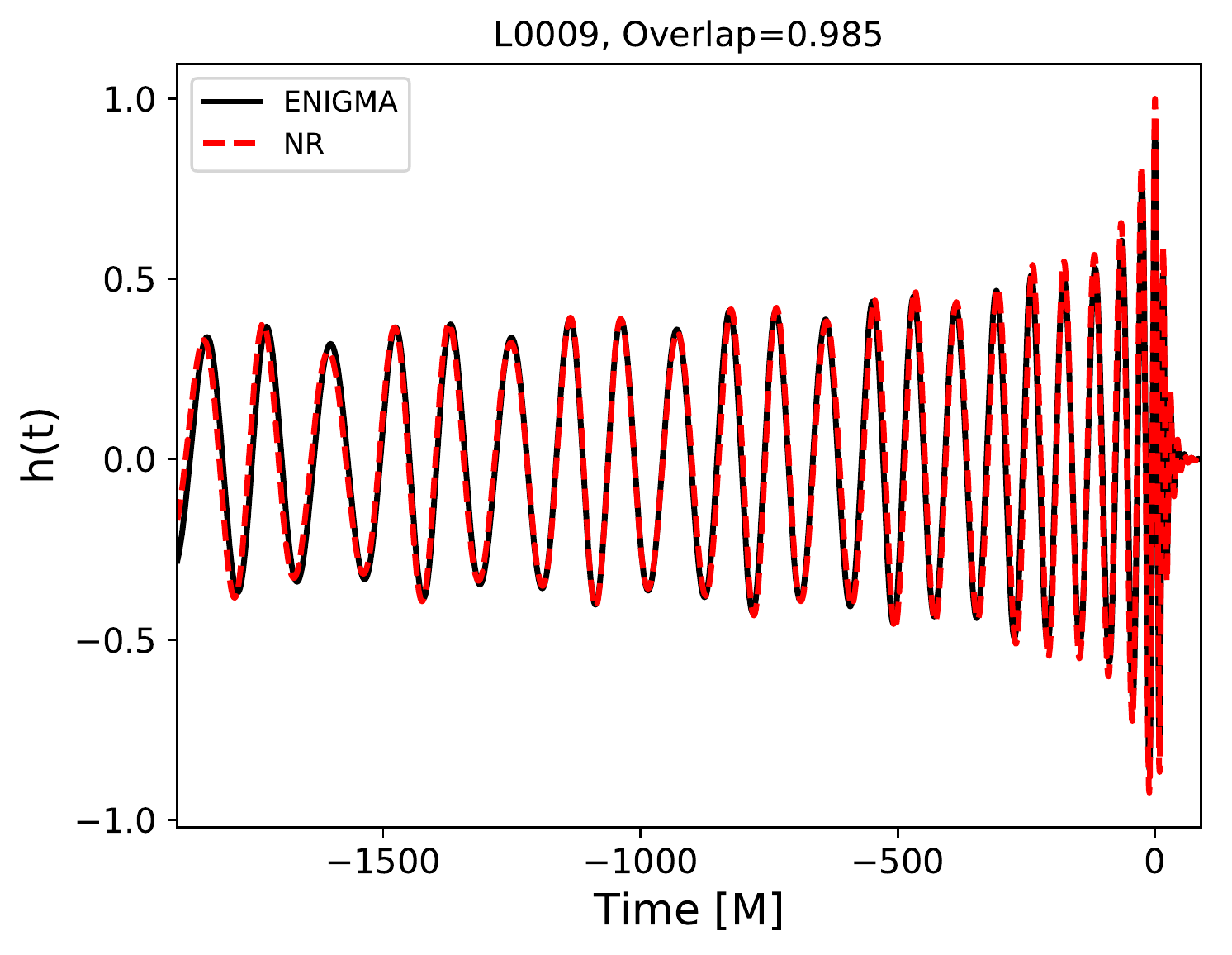}
}
\centerline{
\includegraphics[width=0.5\linewidth]{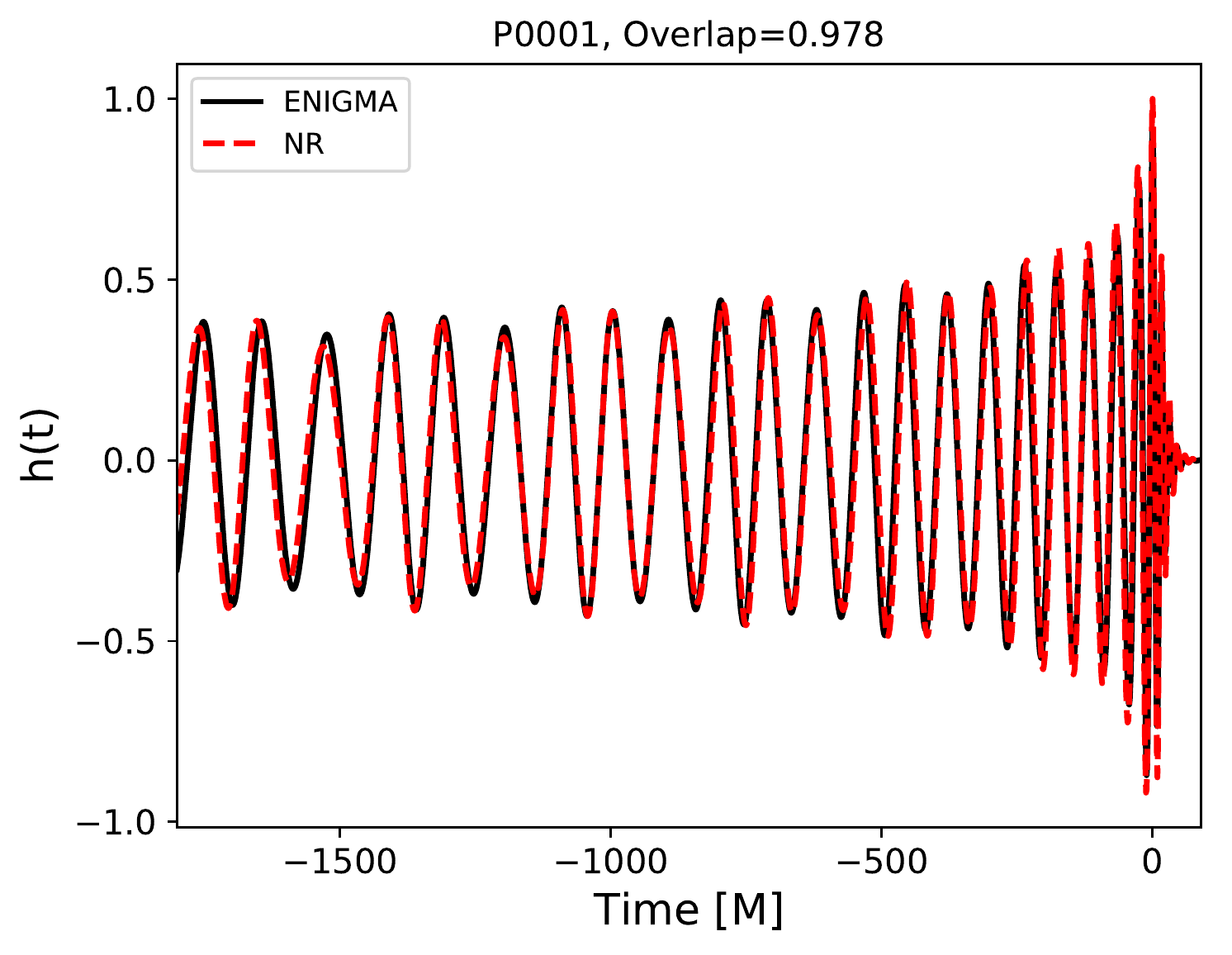}
\includegraphics[width=0.5\linewidth]{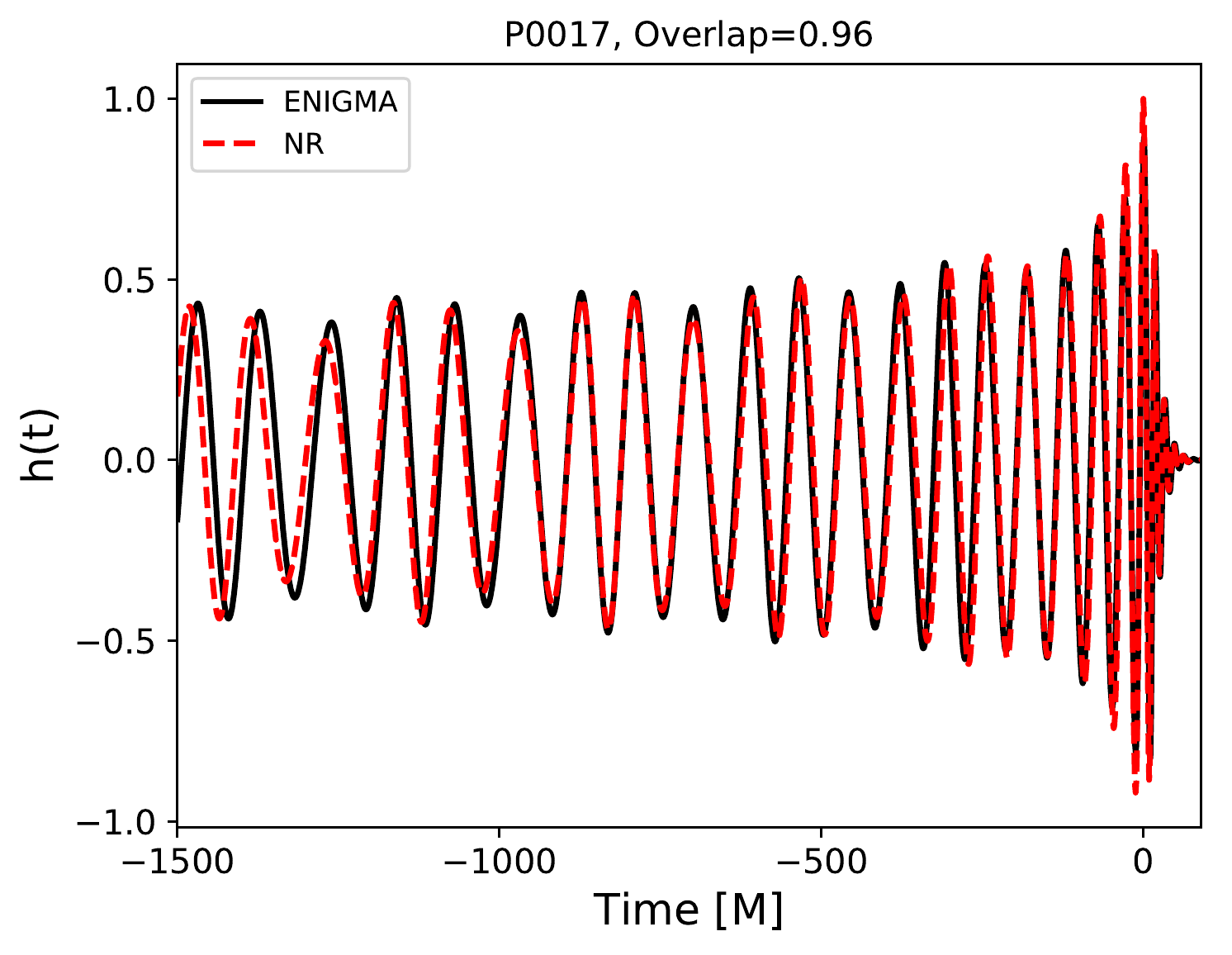}
}
\centerline{
\includegraphics[width=0.5\linewidth]{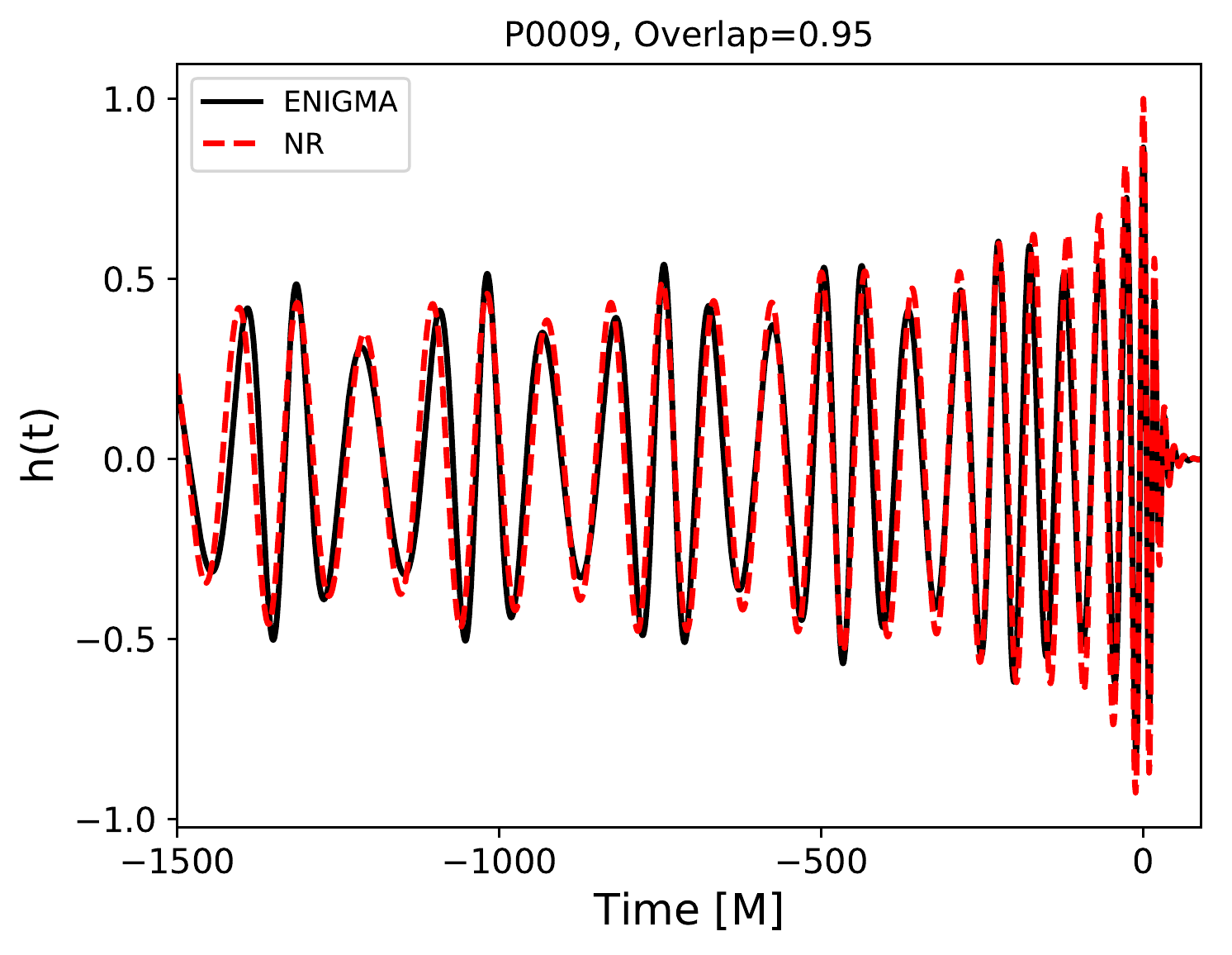}
}
\caption{Comparison between NR waveforms and their \texttt{ENIGMA} counterparts. The \texttt{ENIGMA} waveforms were constructed using the optimal values for the triplet \((\hat{e}_0,\hat{f}_0,\hat{l}_0)\) as determined by the algorithm introduced in Section~\ref{sec:method}.} 
 \label{fig:overlaps_EN_NR}
 \end{figure*}

To demonstrate that our optimization algorithm provides an accurate characterization of eccentric NR waveforms, 
in Figure~\ref{fig:overlaps_EN_NR} we present overlap calculations between \texttt{ENIGMA} waveforms 
using the corresponding optimal triplet \((\hat{e}_0,\, \hat{l}_0, \, \hat{x}_0)\) and their NR counterparts. We note that the overlap between the 
two classes of waveforms is \({\cal{O}}\geq 0.95\). An important finding of these results is that \texttt{ENIGMA} is correctly characterizing NR waveforms of moderately eccentric BBH mergers that have rather asymmetric mass-ratios. 

Given that \texttt{ENIGMA} has only been 
validated, as opposed to calibrated, with NR waveforms, these results indicate that it is possible to capture 
the physics of eccentric compact binary systems by combining in a consistent manner results from several 
analytical relativity approaches, such as PN and BH perturbation theory and the self-force program. Much work still remains to be 
done to provide a robust framework to extend these formalisms to accurately describe the inspiral-merger-ringdown of highly eccentric systems.

As we mentioned above, \texttt{ENIGMA} was constructed under the assumption that moderately eccentric BBH systems circularize prior to merger. When we constructed \texttt{ENIGMA}, we determined the transition point between inspiral and merger by constraining the time window before merger within which state-of-the-art, inspiral-merger-ringdown quasi-circular waveforms~\cite{Bohe:2016gbl} and quasi-circular \texttt{ENIGMA} waveforms have overlaps \({\cal{O}}\geq 0.99\). From this time window, we selected the attachment time closest to merger. At the time of that study we had produced eccentric NR waveforms with mass-ratios \(q\leq5.5\), which we used to validate this approach~\cite{huerta:2018PhRvD}. Having constructed an NR waveform catalog that now covers a deeper parameter space~\cite{ecc_catalog}, we can actually quantify the circularization rate of more asymmetric mass-ratio BBHs that retain non-negligible eccentricities just a few cycles before merger. This is discussed in the following section.


\section{Circularization of eccentric binary black hole systems}
\label{sec:circular}

To quantify the circularization of moderately eccentric BBH mergers, we compare them 
directly to quasi-circular BBH mergers that have identical mass-ratios. In practice, and assuming 
a flat power spectral density, we 
compute the inner product between eccentric NR waveforms, \(h(t)=h_{+} - i h_{\times}\), and their quasi-circular counterparts, \(s(t)=s_{+} - i s_{\times}\),  
using the relation

\begin{align}
\label{eq:overlap}
\left(h(t)\, \vert\, s(t)\right) &= \int_{t_0}^{T}\left[h^*(t)\,s(t) + h(t)\,s^*(t)\right]\mathrm{d}t\,,\\
\label{eq:overlap_expand}
\left(h(t)\, \vert\, s(t)\right) &= \int_{t_0}^{T}\left[h_{+}(t)\,s_{+}(t) + h_{\times}(t)\,s_{\times}(t) \right]\mathrm{d}t\,,
\end{align}

\noindent where \(t_0\) is a fiducial time from which we compute the overlap, and \(T\) corresponds to the merger time. We then compute the normalized overlap

\begin{align}
\label{eq:nor_overlap}
\left(\hat{h}(t)\, \vert\, \hat{s}(t)\right) &= \frac{\left(h(t)\, \vert\, s(t)\right)}{\sqrt{\left(h(t)\, \vert\, h(t)\right)\left(s(t)\, \vert\, s(t)\right)}}\,. 
\end{align}

\noindent Finally, the quantity we quote for our results below is the maximized overlap

\begin{align}
\label{eq:max_overlap}
{\cal{O}}\left(h,\,s\right)&=  \underset{t_c\,, \phi_c}{\textrm{max} }\left(\hat{h}(t)\, \vert\, \hat{s}(t)\right)\,,
\end{align}

\noindent which is obtained by maximizing the normalized overlap over time and phase of coalescence, \((t_c\,, \phi_c)\), respectively.

\begin{figure*}
\centerline{
\includegraphics[width=0.54\linewidth]{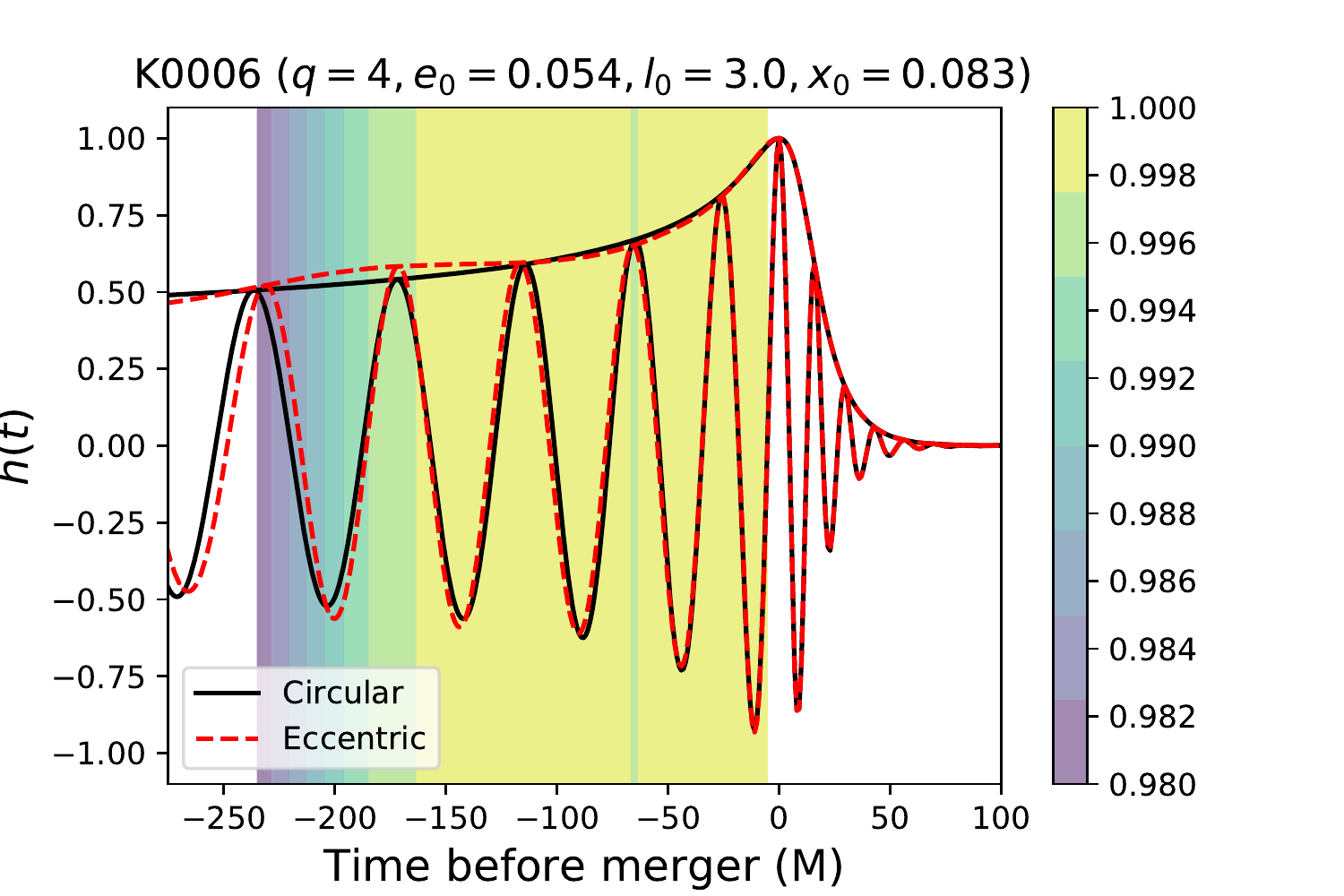}\hspace{-1.9em}%
\includegraphics[width=0.54\linewidth]{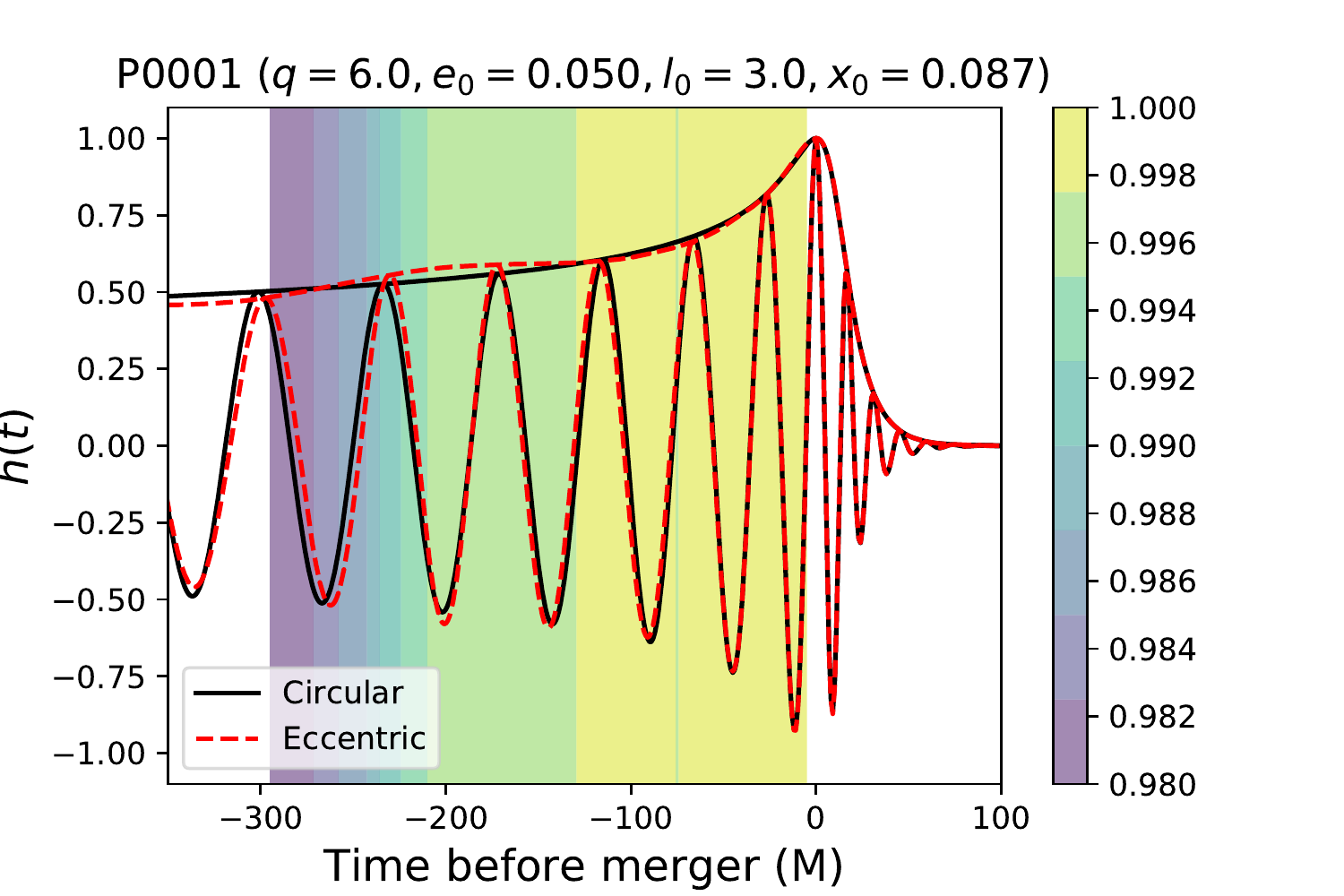}
}
\centerline{
\includegraphics[width=0.54\linewidth]{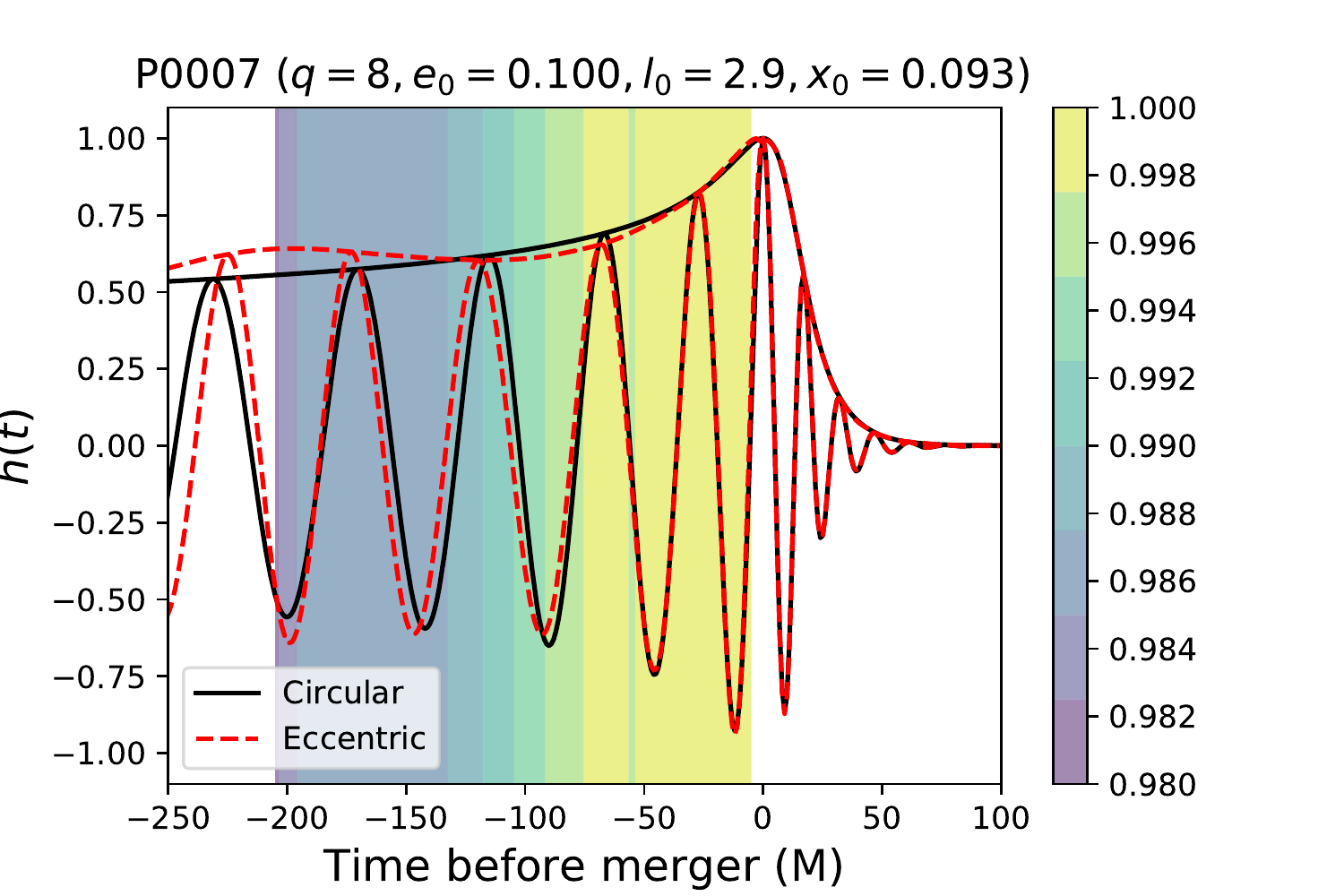}\hspace{-1.9em}%
\includegraphics[width=0.54\linewidth]{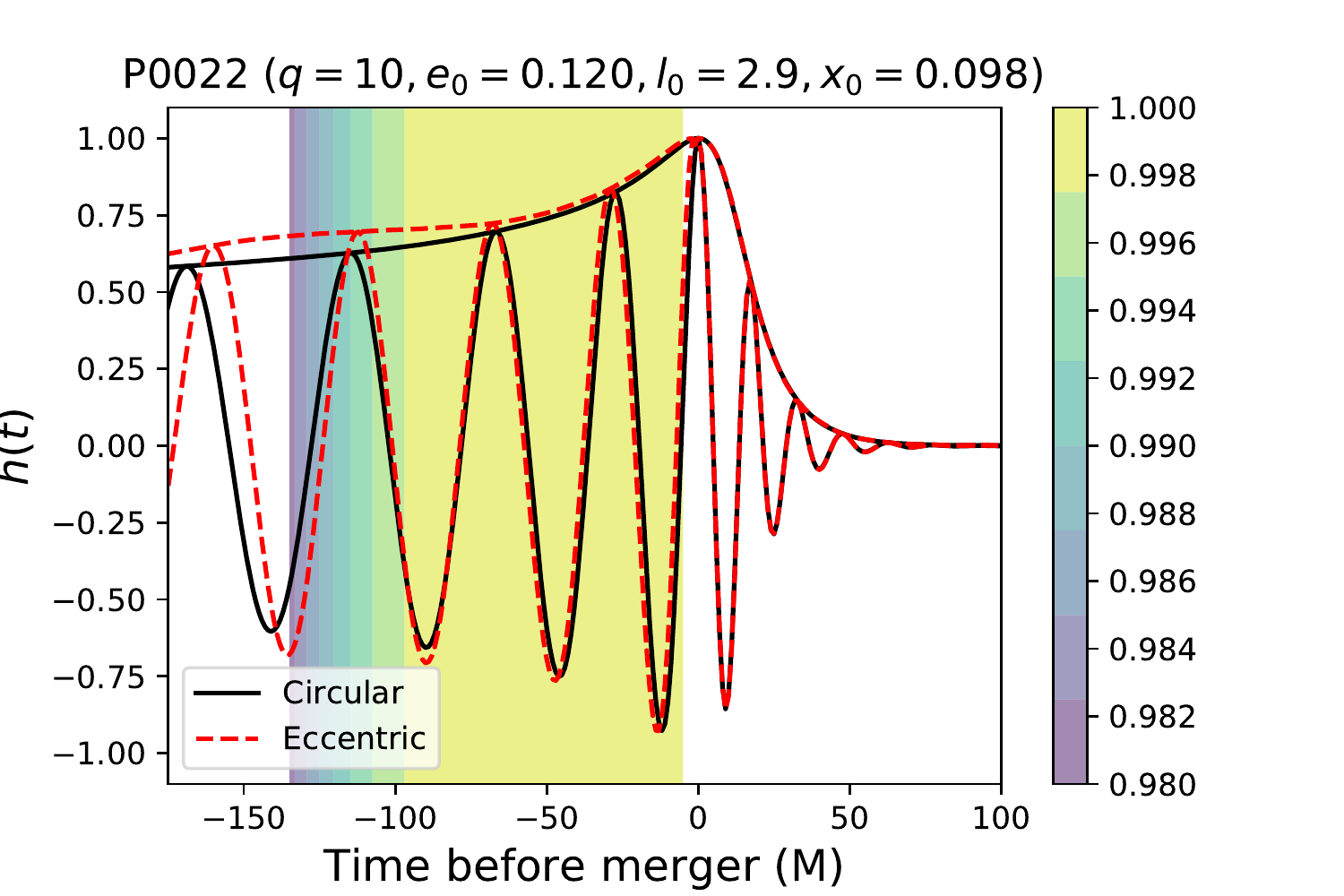}
}
\caption{Overlap between eccentric numerical relativity waveforms and their quasi-circular counterparts. 
We show the overlap, computed between a fiducial time, \(t_0\), and merger, \(t=0M\). These results show that 
moderately eccentric systems circularize prior to merger. Note also that the decrease in overlap as we go 
backwards in time is not monotonically decreasing. This is expected, since the overlap between the quasi-circular and eccentric systems will be different if 
we start the comparison when the eccentric system is close to apoapse or periapse. This distinct property between eccentric and quasi-circular orbits is not 
washed away by maximizing over phase and time of coalescence when computing the overlap.} 
 \label{fig:overlaps}
 \end{figure*}

\noindent We have selected a variety of scenarios to illustrate how mass-ratio and initial eccentricity, \(e_0\), drive the circularization of eccentric BBHs. Figure~\ref{fig:overlaps} indicates that for the more eccentric systems (see Table~\ref{sims}), circularization only happens about \(50M\) before merger. We also notice that the increase in overlap, as \(t_0\rightarrow0\) in Eq.~\eqref{eq:overlap_expand}, is not monotonic. Rather, it has an oscillatory behavior that tracks the eccentric trajectory of the BBH system, and which is clearly captured by the waveform amplitude. We have included the waveform amplitude of the eccentric and quasi-circular signals in the panels of Figure~\ref{fig:overlaps} to clearly show this finding. Notice that as soon as the waveform amplitude of the eccentric signal becomes increasingly monotonic, so does the overlap.

\noindent In view of this analysis, we conclude that modeling eccentric BBH mergers under the assumption that they circularize 
prior to merger requires two key components: (i) an effective scheme that describes the early inspiral evolution, and which 
remains accurate at least \(50M\) before merger; (ii) a stand-alone quasi-circular merger that can be smoothly attached to the late-inspiral evolution. To accomplish this, it is key that future models go beyond low-order PN approximation to describe the radiative and conservative pieces of the waveform dynamics~\cite{Huerta:2017a,huerta:2018PhRvD,moore_2019M,loutrel_lie_2019CQG}.


\section{Higher-order waveform modes}
\label{sec:hom}

The impact of higher-order waveform modes for GW detection in terms of signal-to-noise (SNR) calculations 
has been explored in~\cite{Adam:2018arXiv}. In this section we now quantify the impact of including the modes 
\((\ell, \, \abs{m})= \{(2,\,2),\, (2,\,1),\, (3,\,3),\, (3,\,2), \, (3,\,1),\, (4,\,4),\, (4,\,3),\, (4,\,2)\), \((4,\,1)\}\) in terms 
of overlap calculations. In practice, what we do is to construct two types of signals, one that includes the \(\ell=\abs{m}=2\) mode only, and one that includes all the \((\ell, \, \abs{m})\) modes listed above using the relation

\begin{equation}
h(t,\, \theta,\,\phi) = h_{+} - \textrm{i} h_{\times} = \sum_{\ell\geq2}\,\,\sum_{m \geq -\ell}^{m \leq \ell}h^{\ell m}{}_{-2}Y_{\ell m} \left(\theta, \phi\right)\,, 
\label{strain_ho_modes}
\end{equation} 

\noindent where \({}_{-2}Y_{\ell m} \left(\theta, \phi\right)\) are the spin-weight--2 spherical harmonics~\cite{Blanchet:2006}, and the reference frame described by \((\theta\,,\phi)\), fixed at the center of mass of the BBH, determines the location of the GW detector.
We construct higher-order mode NR waveforms using Eq.~\eqref{strain_ho_modes}, and follow ~\cite{Adam:2018arXiv} to determine the \((\hat{\theta}\,,\hat{\phi})\) combinations that maximize the contribution of \((\ell, \, \abs{m})\) modes for GW detection. In practice, we densely sampled the \((\theta\,,\phi)\) parameter space, and constrained the regions where the integrated amplitude of the \((\ell, \, \abs{m})\)  waveforms is larger than that of their \(\ell=\abs{m}=2\) counterparts. 

\begin{figure*}
\centerline{
\includegraphics[width=0.54\linewidth]{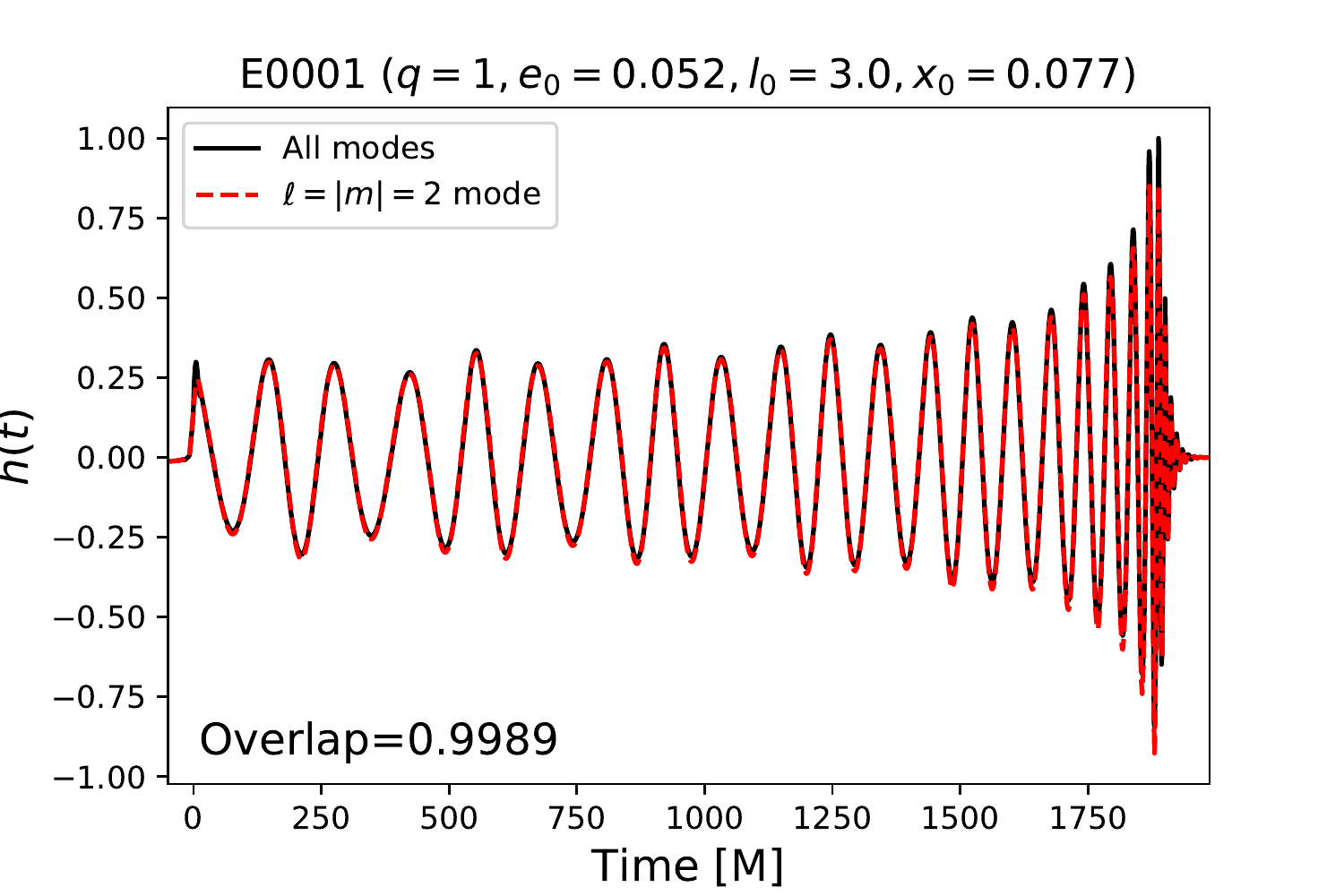}\hspace{-1.9em}%
\includegraphics[width=0.54\linewidth]{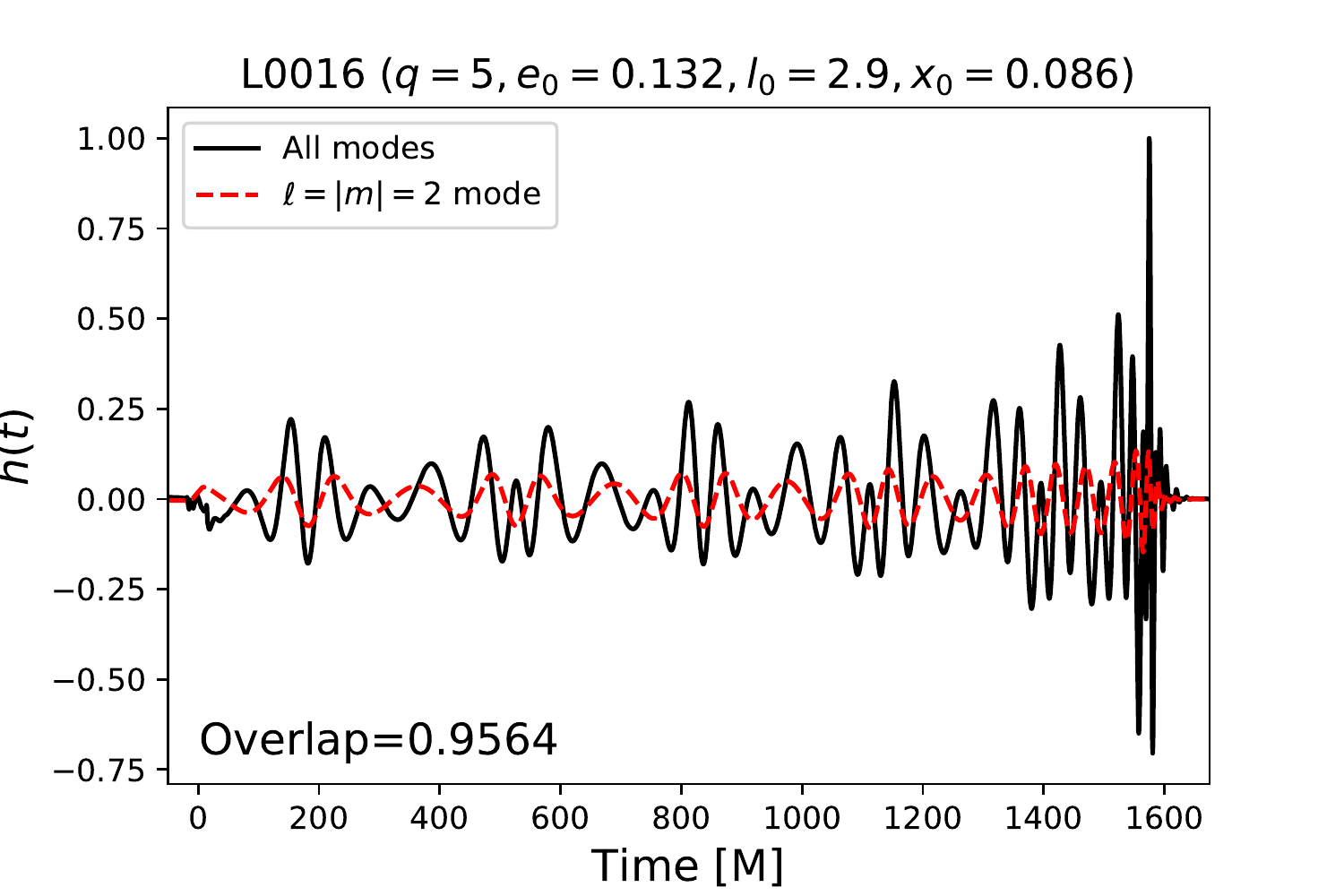}
}
\centerline{
\includegraphics[width=0.54\linewidth]{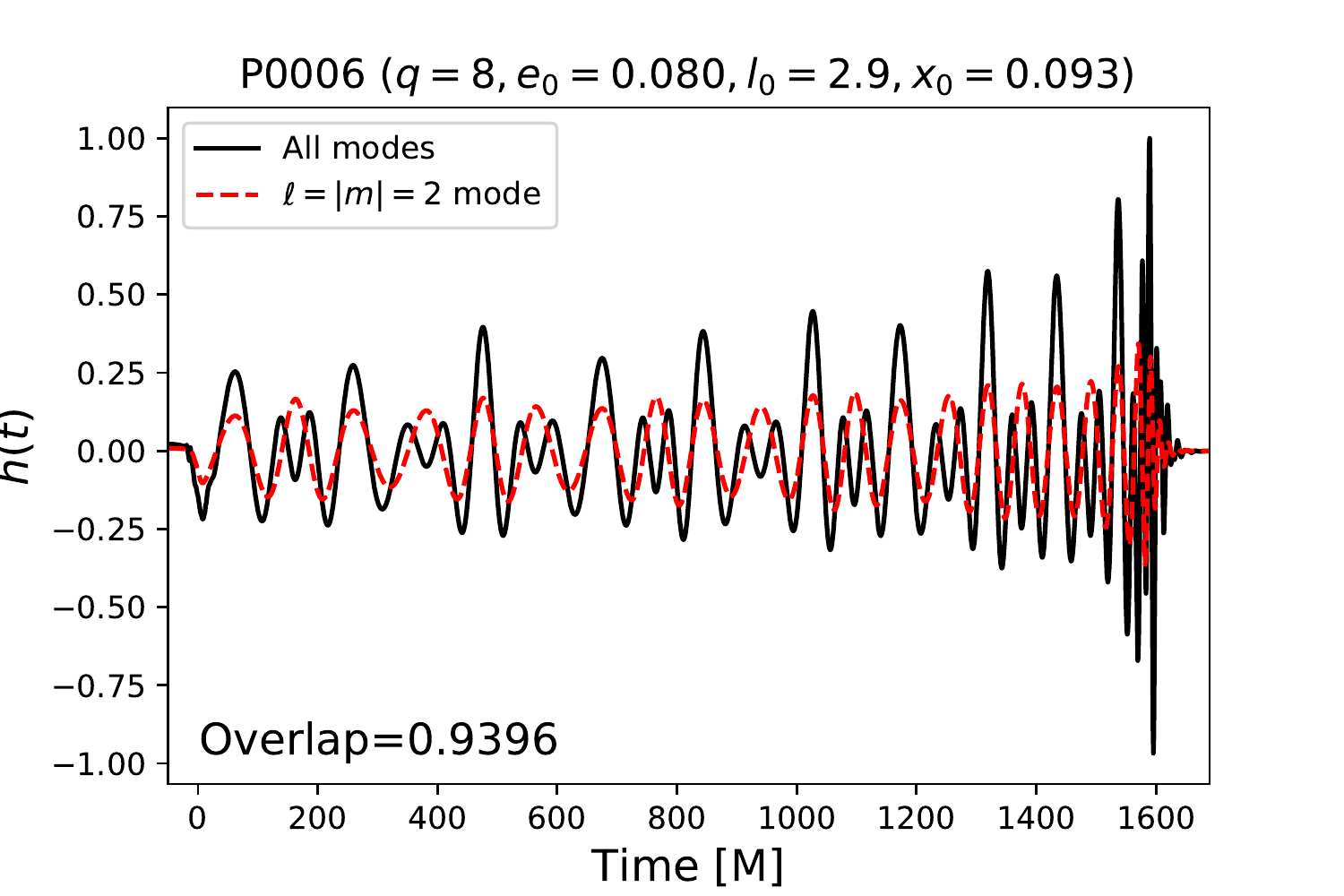}
\includegraphics[width=0.54\linewidth]{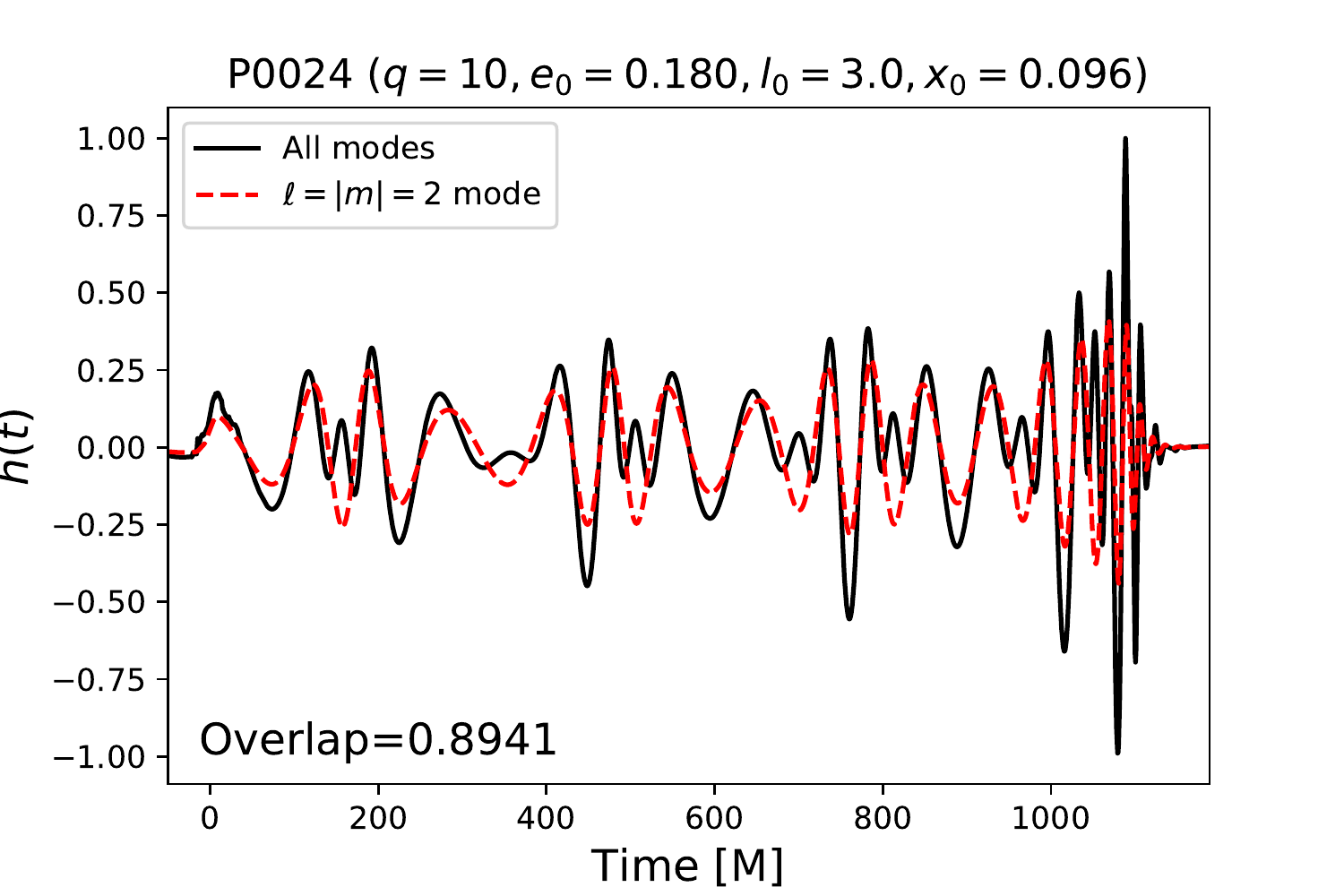}
}
\caption{Overlap between numerical relativity waveforms that include either the modes \((\ell, \, \abs{m})= \{(2,\,2),\, (2,\,1),\, (3,\,3),\, (3,\,2), \, (3,\,1),\, (4,\,4),\, (4,\,3),\, (4,\,2)\), \((4,\,1)\}\) (labelled as ``All modes" in the panels), or just the \((\ell=\abs{m}=2)\) mode. Higher-order waveform models become significant for asymmetric mass-ratio binary black hole systems.}
 \label{fig:overlaps_ho_modes}
 \end{figure*}

Upon constructing NR waveforms with the optimal \((\hat{\theta}\,,\hat{\phi})\) combinations, we compute their overlap using Eq.~\eqref{eq:max_overlap}, but now setting \(t_0\) to the time at which the NR waveform is free from junk radiation, and \(T\) to the final time sample of the NR waveform (see Eq.~\eqref{eq:overlap}). We present results for these calculations for a variety of astrophysically motivated scenarios in Figure~\ref{fig:overlaps_ho_modes}. These results indicate that the inclusion of higher-order modes does not quantitatively modify the morphology of \(\ell=\abs{m}=2\) NR waveforms that describe equal-mass, eccentric BBH mergers. However, NR waveforms that describe asymmetric mass-ratios, eccentric BBH mergers have a much richer topology that requires the inclusion of \((\ell, \, \abs{m})\). 

\noindent Note that the overlap results shown in the four panels in Figure~\ref{fig:overlaps_ho_modes} have been produced using NR waveforms of different lengths. The key point to extract from these analyses is that searches for eccentric BBH mergers that have asymmetric mass-ratios will require signal-processing tools that include higher-order waveform modes. The construction of such algorithms must be pursued in the near future.

\section{Conclusions} 
 \label{sec:end}

In this article we introduced a method to characterize numerical relativity waveforms that describe non-spinning black holes 
on moderately eccentric orbits. To do this, we construct a catalog of \texttt{ENIGMA} waveforms, and then we sift 
through them until we find the optimal combination of parameters that produce an \texttt{ENIGMA} waveform that best 
replicates a given NR waveform. Through this procedure, we optimize three orbital parameters, namely, initial orbital eccentricity, initial mean anomaly, and initial dimensionless orbital frequency. We have demonstrated that when we compute the overlap between our optimized \texttt{ENIGMA} waveforms and their NR counterparts, we obtain overlaps \({\cal{O}}\geq0.95\).

We also quantified the circularization rate of \(\ell=\abs{m}=2\) eccentric NR waveforms by computing the overlap between 
these signals and their quasi-circular counterparts. By choosing a variety of representative systems from our NR catalog~\cite{ecc_catalog}, we found that all NR waveforms, with eccentricities \(e_0\leq0.2\) fifteen cycles before merger, circularize at least \(50M\) before merger. These findings have a variety of implications for ongoing source modeling efforts. For instance, modeling eccentric BBH mergers under the assumption of circularization prior to merger would require a scheme that accurately describes the effects of eccentricity during the early inspiral evolution, and which also remains accurate deep into the strong-field regime up to just a few tens of \(M\) before merger. We expect that ongoing developments in PN theory and in the self-force program will provide the required elements to further enhance the accuracy of existing waveform models to accomplish this goal.

Finally, we explored the need to include higher-order waveform modes to accurately describe the waveform morphology of eccentric BBH mergers. In previous studies, we quantified the importance of \((\ell, \, \abs{m})\) modes for GW detection in terms of SNR calculations. In this study we have broadened that initial approach, showing that the inclusion of \((\ell, \, \abs{m})\) modes is essential for an accurate description of asymmetric mass-ratio, eccentric BBH mergers. 

Having completed these studies, it is now in order to start developing in earnest NR waveform catalogs that describe spinning BBHs on eccentric orbits, and assess the interplay of spin and eccentricity in the dynamical evolution of these GW sources. Extracting observable signatures from their NR waveforms will inform future GW searches that may confirm or rule out the existence of these type of compact binary populations in dense stellar environments. 

\section{Acknowledgements} 
This research is part of the Blue Waters sustained-petascale computing project, which is supported by the National Science Foundation (awards OCI-0725070 and ACI-1238993) and the State of Illinois. Blue Waters is a joint effort of the University of Illinois at Urbana-Champaign and its National Center for Supercomputing Applications. We acknowledge support from the NCSA and the SPIN Program at NCSA. We thank the \href{http://gravity.ncsa.illinois.edu}{NCSA Gravity Group} for useful feedback. NSF-1550514, NSF-1659702, NSF-OAC1659702, and TG-PHY160053 grants are gratefully acknowledged.

\clearpage

\bibliography{references,red_prd,dl_references}
\bibliographystyle{apsrev4-1}

\end{document}